\documentclass[11pt]{article}
 \usepackage{graphicx,amsfonts,amsmath}

 \renewcommand{\title}[1] {%
 \begingroup\begin{center}\vspace{0.0cm}\bf\Large
 \addtolength{\baselineskip}{1mm} #1 \end{center}\endgroup}

 \renewcommand{\author}[1] {%
 \begingroup\begin{center}\vspace{0.2cm}\bf #1 \vspace{0.2cm}
 \end{center}\endgroup}

 \newcommand{\address}[1] {%
 \begingroup\begin{center} #1 \end{center}\endgroup}

\usepackage{graphicx,amsfonts,amsmath,amssymb,latexsym}
\usepackage[english]{babel}
\numberwithin{equation}{section}

\newtheorem{thm}{Theorem}[section]
\newtheorem{defin}[thm]{Definition}
\newtheorem{lemma}[thm]{Lemma}
\newtheorem{prop}[thm]{Proposition}

\newcommand\bz{\bar{z}}
\newcommand\Zb{\mathbb{Z}}
\newcommand\Rb{\mathbb{R}}
\newcommand\Cb{\mathbb{C}}
\newcommand\Pb{\mathbb{P}}
\newcommand\Cy{\mathcal{C}}
\newcommand\ben{\begin{equation*}}
\newcommand\ebn{\end{equation*}}
\newcommand\be{\begin{equation}}
\newcommand\eb{\end{equation}}

\begin{document}
  \title{Tau functions for the Dirac operator on the cylinder}
 \author{O. Lisovyy$^{\;*,\;\dag}$}
 \address{
  $^{*\;}$Bogolyubov Institute for Theoretical Physics \\
  Metrolohichna str., 14-b, Kyiv-143, 03143, Ukraine \vspace{0.2cm} \\
   $^{\dag\;}$D\'epartement de Math\'ematiques, Universit\'e d'Angers,
  \\ 2 Boulevard Lavoisier, 49045 Angers, France}
  \date{}

\begin{abstract}
 The goal of the present paper is to calculate the determinant of the
 Dirac operator with a mass in the cylindrical geometry. The domain of this operator
 consists of functions that realize a unitary one-dimensional
 representation of the fundamental group of the cylinder with $n$
 marked points. The determinant represents a version of the
 isomonodromic $\tau$-function, itroduced by M.~Sato, T.~Miwa and
 M.~Jimbo. It is calculated by comparison of  two sections of the
 $\mathrm{det}^*$-bundle over an infinite-dimensional
 grassmannian. The latter is composed of the spaces of boundary values
 of some local solutions to Dirac equation. The principal ingredients
 of the computation are the formulae for the Green function of the
 singular Dirac operator and for the so-called canonical basis of
 global solutions on the 1-punctured cylinder.  We also derive a set
 of deformation equations satisfied by the expansion coefficients of
 the canonical basis in the general case and find a more explicit
 expression for the $\tau$-function in the simplest case $n=2$.
\end{abstract}
 \section{Introduction}
 The main objective of quantum field theory is the calculation of
 correlation functions of local operators, usually represented via
 functional integrals
 \ben
 \langle\mathcal{O}_1(x_1)\ldots\mathcal{O}_n(x_n)\rangle=
 \frac{\int\mathcal{D}\varphi\;\mathcal{O}_1(x_1)\ldots\mathcal{O}_n(x_n)e^{S[\varphi]}}
 {\int\mathcal{D}\varphi\;e^{S[\varphi]}}\; .
 \ebn
 For a generic interacting QFT such calculation can be done only
 by means of perturbation theory. However, in two dimensions there is
 an interesting way to construct an interacting theory from the free
 one. Let us consider the action of free massive Dirac fermions in
 the flat spacetime,
 \ben
 S[\psi,\bar{\psi}]=\frac12\int d^2x \bar{\psi}D\psi.
 \ebn
 Correlation functions of the (interacting) monodromy fields are defined as
 \be\label{cf}
 \langle\mathcal{O}^{\lambda_1}(a_1)\ldots\mathcal{O}^{\lambda_n}(a_n)\rangle=
 \frac{\int\mathcal{D}\psi\mathcal{D}\bar{\psi}\; e^{{\frac12\int d^2x \;\bar{\psi}D^{a,\lambda}\!\psi}}}
 {\int\mathcal{D}\psi\mathcal{D}\bar{\psi}\; e^{{\frac12\int d^2x\; \bar{\psi}D\psi}}}\; .
 \eb
 The integration in the numerator is performed over field
 configurations that are branched at the points $a_{\nu}$
 with the monodromies $e^{2\pi i \lambda_{\nu}}$ ($\nu=1,\ldots,n$). This change of domain of the Dirac operator
 is symbolically reflected by indexing $D^{a,\lambda}$. The integrals in
 (\ref{cf}) can be formally evaluated to the determinants of the
 corresponding operators,
 \be\label{cf2}
 \langle\mathcal{O}^{\lambda_1}(a_1)\ldots\mathcal{O}^{\lambda_n}(a_n)\rangle=
 \frac{\mathrm{det}\, D^{a,\lambda}}
 {\mathrm{det}\, D\;\;\,\;}\; .
 \eb
 Note, however, that the RHSs of both (\ref{cf}) and (\ref{cf2}) are equally
 ill-defined quantities.

 The determinants of Dirac operators on compact manifolds are usually
 determined via the $\zeta$-function regularization. Starting from
 \cite{ray}, they have been
 extensively studied in the mathematical literature.
 The massless Dirac operators on Riemann surfaces deserved
 special attention, as multiloop contributions to the partition function
 in the string theory are expressed through their determinants
(rigorously defined by D.~Quillen in \cite{quillen}). These
determinants can be thought of as the functions on the moduli space of
 complex structures on the surface.

 In the case we are interested in the Dirac operator is defined not on
 a compact manifold, but on the universal covering of a surface with
 marked points, the determinant being a function of their
 positions.
The problem of rigorous definition of the determinant and the Green function
 for the Dirac operator with branching points on the Euclidean plane
 was solved by Palmer in \cite{pacific}. His work relies heavily on
 the analysis of monodromy preserving deformations for the Dirac operator,
 developed earlier by Sato, Miwa and Jimbo \cite{smj}. More precisely,
 Palmer's determinant represents another version of SMJ
 $\tau$-function. Its logarithmic derivatives with respect to the
 coordinates of branching points are expressed via the expansion
 coefficients of some special solutions to Dirac equation, that can be
 constructed from the so-called canonical basis of solutions. The
 theory of isomonodromic deformations gives a set of nonlinear
 differential equations satisfied by these expansion
 coefficients. Moreover, in the simplest case $n=2$ an explicit
 formula for the determinant was found \cite{pacific}.

 Later similar results were obtained for the Dirac operator on
 the Poincar\'e disk \cite{narayanan,palmer_tracy,beatty}. In this
 connection we should also  mention the recent work of Doyon \cite{doyon},
 where the two-point correlation function of monodromy fields in the
 hyperbolic geometry was calculated by field-theoretic methods.

  In the present paper, we define and calculate the determinant of
  the massive Dirac operator in cylindrical geometry. The latter corresponds to QFT in the
  finite volume or at non-zero temperature. This work was inspired by
  recent progress in the study of Ising model --- calculation of
  finite-size correlation functions \cite{bugrij2,bugrij}, spin matrix
  elements \cite{PhysLetts} and direct derivation of the differential
  equations satisfied by two-point correlator in the continuum limit
  \cite{ya}. Ising model is related to the above theory with special
  monodromy $\lambda_{\nu}=\pm\frac12$.

 This paper is organized as follows.
 In the next section  we introduce the canonical basis of global solutions to the
 Dirac equation on the cylinder and calculate it explicitly for
 $n=1$ (see Theorem~\ref{teo2}). All subsequent  computations are
 based on these formulae. The Green function of the singular Dirac
 operator is defined in Section~3. Its derivatives with respect to the
 coordinates of branching points are expressed through some solutions
 to the Dirac equation and have remarkable factorized form (formulae
 (\ref{der1}) and (\ref{der2})). At the end of the section, the  Green function on the
 1-punctured cylinder is computed (see (\ref{eq34})-(\ref{eq35})). Section~4 is devoted to the
 definition and calculation of the $\tau$-function. We introduce the
 $\mathrm{det}^*$-bundle over an infinite-dimensional grassmannian
 that consists of the spaces of boundary values of some local
 solutions to the Dirac equation. The $\tau$-function is obtained by
 comparison of the canonical section of this $\mathrm{det}^*$-bundle
 with a section that is constructed using the one-point Green
 functions. The logarithmic derivatives of the $\tau$-function
 can  also be written in terms of the expansion coefficients of the solutions
 (\ref{basis2}). This shows that it is independent of the chosen
 localization. As an illustration, we find a more explicit expression for
 the $\tau$-function when $n=2$ (formula (\ref{taun2})).
  Finally, in the Section~5 a set of deformation
 equations for the expansion coefficients is derived. We
 conclude with a brief discussion of possible generalizations, open
 problems and application of obtained results in the quantum field theory
 at non-zero temperature.

 \section{Canonical basis of solutions to Dirac equation}
 \subsection{Definitions}
 Let $a=(a_1,\ldots, a_n)$ be a collection of $n$ distinct points on the
 cylinder $\Cy$. The fundamental group $\pi_1(\Cy\backslash a;x_0)$ is generated
 by  homotopy classes of $n+1$ loops
 $\gamma_0,\ldots,\gamma_n$ shown in the Fig.~1. It acts on
 the universal covering $\widetilde{\Cy\backslash a}$ by deck
 transformations. Let us fix a one-dimensional unitary representation
 \be\label{repr}
 \rho_{\lambda}:\;\pi_1(\Cy\backslash a;x_0)\rightarrow U(1), \qquad
 [\gamma_{\nu}]\mapsto e^{-2\pi i \lambda_{\nu}},
 \qquad \nu=0,\ldots,n,
 \eb
 \ben
 \lambda_0\in\Rb,\qquad\lambda_{\nu}\in \Rb\backslash\Zb, \qquad \nu=1,\ldots,n.
 \ebn
 \begin{figure}[h]
\begin{center}
 \includegraphics[height=3.5cm]{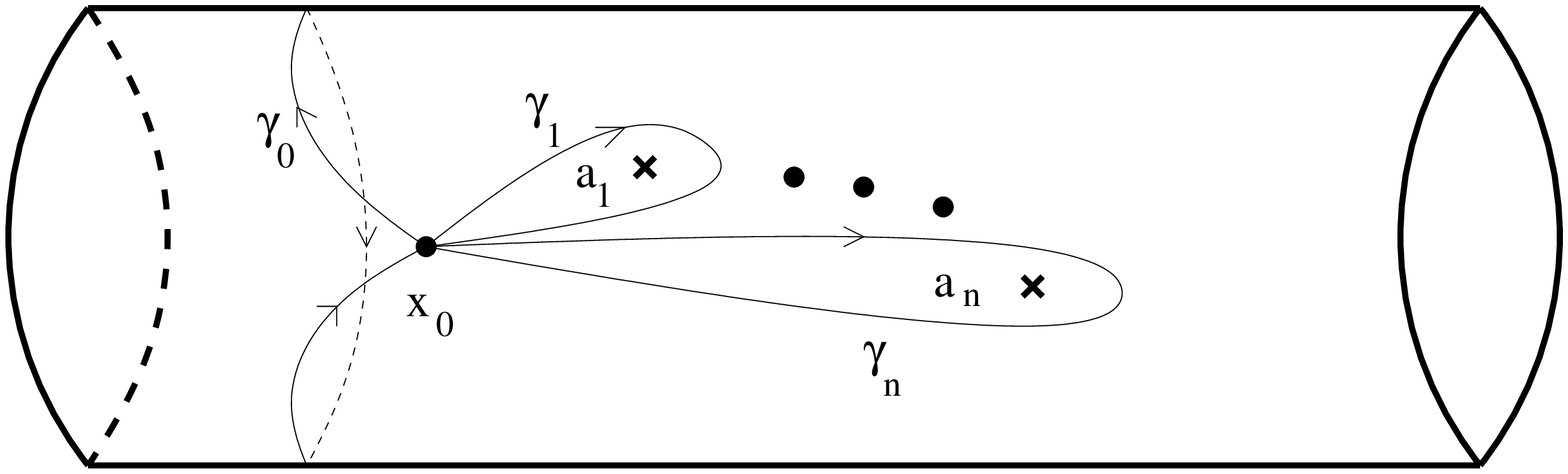}\vspace{0.5cm}

 Fig. 1.
 \end{center}
 \end{figure}

 As usual, we replace the cylinder by the strip $S=\{(x,y)\in
 \Rb^2:0\leq y\leq \beta\}$
 whose upper and lower edges are identified. The Dirac
 operator on $\widetilde{\Cy\backslash a}$ is induced by the Dirac operator on
 $\Rb^2$, which can be written as
 \be\label{operator}
 D=\left(\begin{array}{cr}
 \frac m2 & -\partial_z \\
 -\partial_{\bz} & \frac m2 \
 \end{array}\right),
 \eb
 where $z$, $\bz$ --- standard complex coordinates
 \ben
\begin{cases}
 z=x+iy, \\
 \bz=x-iy,
\end{cases}
\qquad
\begin{cases}
 \partial_z=\frac12\,(\partial_x-i\partial_y), \\
 \partial_{\bz}=\frac12\,(\partial_x+i\partial_y).
\end{cases}
 \ebn
 We are looking for multivalued solutions
 $\tilde{\psi}:\widetilde{\Cy\backslash a}\rightarrow\Cb^2$ to Dirac equation that transform
 according to the representation (\ref{repr}),
 $$ D\tilde{\psi}(x)=0,\qquad \tilde{\psi}(\gamma
 x)=\rho_{\lambda}([\gamma])\cdot\tilde{\psi}(x).$$

 This problem can be reformulated as follows. Fix a system of
 branchcuts $b=(b_1,\ldots,b_n;$ $ d_0,\ldots,d_n)$ shown in the Fig.~2 and consider the
 solutions to
 Dirac equation on  $\Cy\backslash b$ that can be continued across the
 branchcuts away from the points $a_1,\ldots,a_n$. The solutions we
 are interested in have left and right continuations across $b_{\nu}$
 that differ by the factor $e^{2\pi i \lambda_{\nu}}$
 ($\nu=1,\ldots,n$). The continuations across $d_{\nu}$ differ by
 $\exp\bigl(2\pi i \sum\limits_{k=0}^{\nu}\lambda_{k}\bigr)$, $\nu=0,\ldots,n$.
  \begin{figure}[h]
\begin{center}
 \includegraphics[height=4.5cm]{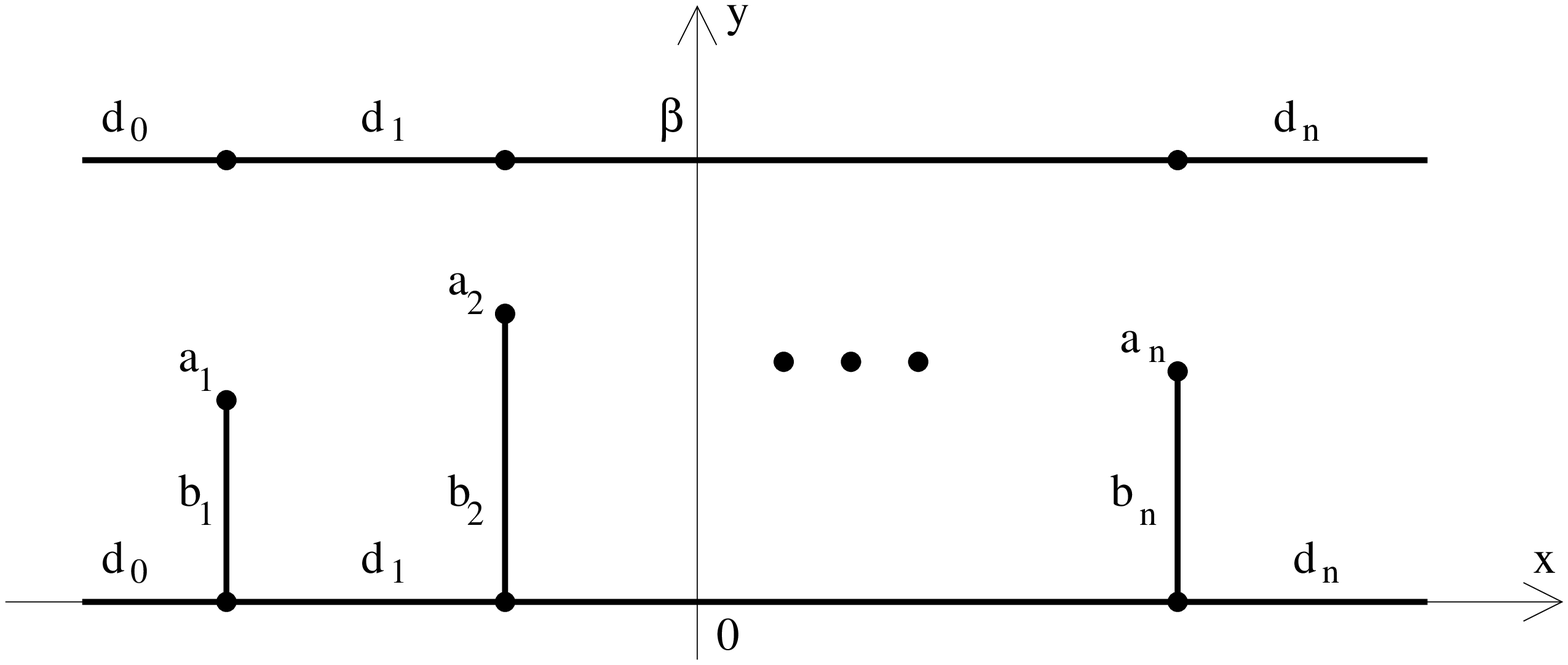}\vspace{0.5cm}

 Fig. 2. $\qquad$
 \end{center}
 \end{figure}

 To describe the local behaviour of such solutions in the neighbourhood of
 the point $a_{\nu}$, consider an open disk $B$ of sufficiently small but
 finite radius, centered at $a_{\nu}$, and introduce in $B$ polar coordinates
 \ben
  \begin{cases}
    r={\left|z-a_{\nu}\right|}^{1/2}, \\
    \varphi=\frac{1}{2i}\ln\frac{z-a_{\nu}}{\bz-\bar{a}_{\nu}}\, ,
  \end{cases}\qquad
  \begin{cases}
 \partial_z=\frac12\,e^{-i\varphi}(\partial_r-\frac{i}{r}\,\partial_{\varphi}), \\
 \partial_{\bz}=\frac12\,e^{i\varphi}(\partial_r+\frac{i}{r}\,\partial_{\varphi}).
\end{cases}
 \ebn
 The local form of the Dirac operator on $B$ is then
 \ben
 D=\frac12\,\left(\begin{array}{cc}
  m & -e^{-i\varphi}\bigl(\partial_r-\frac{i}{r}\,\partial_{\varphi}\bigr) \\
 -e^{i\varphi}\bigl(\partial_r+\frac{i}{r}\,\partial_{\varphi}\bigr) &  m \
 \end{array}\right).
 \ebn
 Since for any multivalued solution $\psi$ the function $e^{-
   i\lambda_{\nu}\varphi}\psi$ is single-valued on $B$, it can be expanded in
 Fourier series. Substituting the series into Dirac equation, one
 obtains \cite{smj}
 \be\label{fourier}
 \psi[a_{\nu}]=\sum\limits_{k\in \Zb+\frac12}\left\{a_k w_{k+\lambda_{\nu}}[a_{\nu}]+
 b_k w_{k-\lambda_{\nu}}^{\;*}[a_{\nu}]\right\}.
 \eb
 where
 \be\label{ssolutions}
 w_l[a_{\nu}]=\left(\begin{array}{r}
  e^{i(l-1/2)\varphi}I_{l-1/2}(mr) \\
  e^{i(l+1/2)\varphi}I_{l+1/2}(mr)
 \end{array}\right),\qquad
 w_l^*[a_{\nu}]=\left(\begin{array}{c}
 e^{-i(l+1/2)\varphi}I_{l+1/2}(mr) \\
  e^{-i(l-1/2)\varphi}I_{l-1/2}(mr)
 \end{array}\right),
 \eb
 and $I_l(x)$ is the modified Bessel function of the first kind.

 To obtain some kind of regularity, we put certain conditions on the singular behaviour of the
 function $\psi$ at the point $a_{\nu}$. There are two essential types of constraints:
 \begin{itemize}
 \item Let $0<\lambda_{\nu}<1$ and require $\psi$ to be square
 integrable in the neighborhood of $a_{\nu}$. When $|z|\rightarrow a_{\nu}$,
 the asymptotics of special solutions  has the form
 \ben
 w_l[a_{\nu}]\sim\left(\begin{array}{r}
  \frac{(m(z-a_{\nu})/2)^{l-\frac12}}{(l-\frac12)!} \\
  \frac{(m(z-a_{\nu})/2)^{l+\frac12}}{(l+\frac12)!}
 \end{array}\right)+\ldots,\qquad
 w_l^*[a_{\nu}]\sim\left(\begin{array}{c}
  \frac{(m(\bz-\bar{a}_{\nu})/2)^{l+\frac12}} {(l+\frac12)!}\\
  \frac{(m(\bz-\bar{a}_{\nu})/2)^{l-\frac12}}{(l-\frac12)!}
 \end{array}\right)+\ldots,
 \ebn
 where factorials are understood as $l!=\Gamma(l+1)$. Then to
 satisfy the condition of square integrability, a part of
 coefficients in (\ref{fourier}) must vanish,
  \be\label{l2}
 \psi[a_{\nu}]=a_{-1/2}w_{-1/2+\lambda_{\nu}}[a_{\nu}]+\sum\limits_{k>0}\left\{a_k w_{k+\lambda_{\nu}}[a_{\nu}]+
 b_k w_{k-\lambda_{\nu}}^{\;*}[a_{\nu}]\right\}.
 \eb
 \item Let $-\frac12<\lambda_{\nu} <\frac12$ and require
 \be\label{bcond}
 \left(\begin{array}{cc}(z-a_{\nu})^{-\lambda_{\nu}}&0\\0&(\bz-\bar{a}_{\nu})^{\lambda_{\nu}}\end{array}\right)\psi\in H^1[a_{\nu}],
 \eb
 where $H^1[a_{\nu}]$ denotes the space of functions that are
 single-valued and square integrable in the neighborhood of
 $a_{\nu}$ together with their first derivatives. Then
   \be\label{h1}
 \psi[a_{\nu}]=\sum\limits_{k>0}\left\{a_k w_{k+\lambda_{\nu}}[a_{\nu}]+
 b_k w_{k-\lambda_{\nu}}^{\;*}[a_{\nu}]\right\}.
 \eb
 \end{itemize}
 Now let us consider multivalued solutions with monodromy
 (\ref{repr}) that are square integrable at $|x|\rightarrow\infty$ and satisfy (\ref{l2}) or
 (\ref{h1}) in the neighborhood of \textit{each} singularity.
 The spaces of solutions of the
 first and second type will be denoted by $\mathbf{W}^{a,\lambda}$ and
 $\widetilde{\mathbf{W}}^{a,\lambda}$ respectively.
 \begin{thm}\label{teo1} $\mathrm{dim}\,\mathbf{W}^{a,\lambda}\leq n$;
 $\mathrm{dim}\,\widetilde{\mathbf{W}}^{a,\lambda}=0$.
 \end{thm}
 $\blacksquare$  Consider a positive definite scalar
 product on $\mathbf{W}^{a,\lambda}$:
 \be\label{product}
 \langle u,w\rangle=\overline{\langle
 w,u\rangle}=\frac{m^2}{2}\int\limits_{\Cy\backslash
 a}\bar{u}\cdot w\,idz\wedge d\bz=\frac{m^2}{2}\int\limits_{\Cy\backslash
 a}(\bar{u}_1 w_1+\bar{u}_2 w_2)\,idz\wedge d\bz  .
 \eb
 Note that the expression under the integral is indeed a single-valued
 function on $\Cy\backslash a$. This function is integrable  due to
 imposed boundary conditions. From the Dirac
 equation on $\Cy\backslash b$ it follows that
 \ben
 \begin{cases}
 \frac m2\,w_1=\partial_z w_2, \\
 \frac m2\,w_2=\partial_{\bz} w_1,
 \end{cases}\qquad
  \begin{cases}
 \frac m2\,\bar{u}_1=\partial_{\bz} \bar{u}_2, \\
 \frac m2\,\bar{u}_2=\partial_z \bar{u}_1,
 \end{cases}
 \ebn
 and we get
 \be\label{stokes1}
 \frac m2 \,(\bar{u}_1 w_1+\bar{u}_2 w_2)\,dz\wedge d\bz=-d(\bar{u}_2 w_1
 dz)=d(\bar{u}_1 w_2 d\bz).
 \eb
 Denote by $D_{\varepsilon}(a_{\nu})$ the disk of radius $\varepsilon$ about
 $a_{\nu}$. Using (\ref{stokes1}) and Stokes theorem, one obtains
 \ben
 \langle u,w\rangle=
 im \sum\limits_{\nu=1}^n \;\lim_{\varepsilon\rightarrow0}
 \oint\limits_{\partial D_{\varepsilon}(a_{\nu})}
 \bar{u}_2w_1 dz=
 \ebn
 \ben
 =im \sum\limits_{\nu=1}^n \;\lim_{\varepsilon\rightarrow0}
 \oint\limits_{\partial D_{\varepsilon}(a_{\nu})}
 \left(\overline{a_{-1/2}^{(\nu)}(u)}\,\frac{(m\bz/2)^{\lambda_{\nu}}}{\lambda_{\nu}!}+
 \ldots+\overline{b_{1/2}^{(\nu)}(u)}\,\frac{(mz/2)^{-\lambda_{\nu}}}{(-\lambda_{\nu})!}+
 \ldots\right)\times
 \ebn
 \ben
 \times\left(a_{-1/2}^{(\nu)}(w)\,\frac{(mz/2)^{\lambda_{\nu}-1}}{(\lambda_{\nu}-1)!}+
 \ldots+b_{1/2}^{(\nu)}(w)\,\frac{(m\bz/2)^{1-\lambda_{\nu}}}{(1-\lambda_{\nu})!}+
 \ldots\right)dz=
 \ebn
 \be\label{product1}
 =-4\sum\limits_{\nu=1}^n
 \overline{b_{1/2}^{(\nu)}(u)}\,a_{-1/2}^{(\nu)}(w)\sin\pi\lambda_{\nu}.
 \eb
 Or, analogously
 \ben
 \langle u,w\rangle=
 -im \sum\limits_{\nu=1}^n \;\lim_{\varepsilon\rightarrow0}
 \oint\limits_{\partial D_{\varepsilon}(a_{\nu})}
 \bar{u}_1w_2 d\bz=
 \ebn
 \ben
 =-im \sum\limits_{\nu=1}^n \;\lim_{\varepsilon\rightarrow0}
 \oint\limits_{\partial D_{\varepsilon}(a_{\nu})}
 \left(\overline{a_{-1/2}^{(\nu)}(u)}\,\frac{(m\bz/2)^{\lambda_{\nu}-1}}{(\lambda_{\nu}-1)!}+
 \ldots+\overline{b_{1/2}^{(\nu)}(u)}\,\frac{(mz/2)^{1-\lambda_{\nu}}}{(1-\lambda_{\nu})!}+
 \ldots\right)\times
 \ebn
 \ben
 \times\left(a_{-1/2}^{(\nu)}(w)\,\frac{(mz/2)^{\lambda_{\nu}}}{\lambda_{\nu}!}+
 \ldots+b_{1/2}^{(\nu)}(w)\,\frac{(m\bz/2)^{-\lambda_{\nu}}}{(-\lambda_{\nu})!}+
 \ldots\right)d\bz=
 \ebn
 \be\label{product2}
 =-4\sum\limits_{\nu=1}^n
 \overline{a_{-1/2}^{(\nu)}(u)}\,b_{1/2}^{(\nu)}(w)\sin\pi\lambda_{\nu}=
 \overline{\langle w,u\rangle}.
 \eb
 If the dimension of $\mathbf{W}^{a,\lambda}$ were greater than
 $n$, we would be able to construct a solution $v\in\mathbf{W}^{a,\lambda}$ with  all
 $a_{-1/2}^{(\nu)}(v)=0$ ($\nu=1,\ldots,n$). This solution
 has zero norm $\langle v,v\rangle=0$ and, therefore,
 vanishes identically, implying the first statement of the
 theorem.

 Note that the solutions of the second type are square integrable
 with respect to the inner product (\ref{product}). We can
 show in absolutely analogous fashion that $\langle v,v\rangle=0$
 for all $v\in\widetilde{\mathbf{W}}^{a,\lambda}$. Consequently,
 $\mathrm{dim}\,\widetilde{\mathbf{W}}^{a,\lambda}=0$. $\blacksquare$

 Suppose\footnote{The proof is based on some technique from functional
   analysis and is very close to the proof of   Theorem~3.2.4 in \cite{smj}.} that $\mathrm{dim}\,\mathbf{W}^{a,\lambda}=n$. Then we
 can fix a \textit{canonical basis} $\{\mathbf{w}_{\mu}\}_{\mu=1,\ldots,n}$ of this
 space, having chosen
 $a_{-1/2}^{(\nu)}(\mathbf{w}_{\mu})=\delta_{\mu\nu}$:
 \be\label{basis}
 \mathbf{w}_{\mu}[a_{\nu}]=\delta_{\mu\nu}w_{-1/2+\lambda_{\nu}}[a_{\nu}]+
 \sum\limits_{k>0}\left\{a_{\;k}^{(\nu)}(\mathbf{w}_{\mu}) w_{k+\lambda_{\nu}}[a_{\nu}]+
 b_{\;k}^{(\nu)}(\mathbf{w}_{\mu}) w_{k-\lambda_{\nu}}^{\;*}[a_{\nu}]\right\}
 \eb
\textbf{Remark}. Let us calculate the inner product of two
 elements of the canonical basis in two ways --- by the formula
 (\ref{product1}) and its ``conjugate'' (\ref{product2}):
 \be\label{prelim}
 \langle\mathbf{w}_{\mu},\mathbf{w}_{\nu}\rangle=
 -4\overline{b_{1/2}^{(\nu)}(\mathbf{w}_{\mu})}\sin\pi\lambda_{\nu}=
 -4b_{1/2}^{(\mu)}(\mathbf{w}_{\nu})\sin \pi\lambda_{\mu}.
 \eb
 We have obtained a set of algebraic relations between the
 expansion coefficients $b_{1/2}^{(\nu)}(\mathbf{w}_{\mu})$.
 In what follows, we will deduce additional relations and use them
 in the construction of deformation equations.\vspace{0.3cm}

 The ``planar'' analog of the previous theorem has an instructive
 illustration when $n=1$. In the case of a single branching point one can
 suppose it to lie at zero. Then any solution with required singular
 behaviour is represented  by the expansion
 $$\psi=a_{-1/2}w_{-1/2+\lambda}[0]+\sum\limits_{k>0}\left\{a_k w_{k+\lambda}[0]+
 b_k w_{k-\lambda}^{\;*}[0]\right\}$$
 on the whole punctured plane $\Rb^2\backslash\{0\}$. This expansion
 will be square integrable at infinity if and only if
 \ben
 \begin{cases}
 a_k=0 \text{ for } k>0, \\
 b_k=0 \text{ for } k>1, \\
 b_{1/2}=-a_{-1/2},
 \end{cases}
 \ebn
 since the only integrable combinations of partial solutions
 (\ref{ssolutions}) are
 \ben
 \hat{w}_{l}[0]=w_{-l}^{\,*}[0]-w_l[0].
 \ebn
 Then, as one could expect, for $n=1$ the space $\mathbf{W}^{0,\lambda}$ is generated by
 the single element of canonical basis
 \ben
 \mathbf{w}=w_{-1/2+\lambda}[0]-w_{1/2-\lambda}[0]=-\hat{w}_{1/2-\lambda}[0].
 \ebn
 With some efforts, it is also possible to find an explicit formula for the
 canonical basis on the 1-punctured cylinder. This problem will be
 solved in the next subsection, using a generalization of the
 method of Fonseca and Zamolodchikov \cite{fonseca}.
 \subsection{Canonical basis on the cylinder with one branching point}
 We are looking for the solution $\psi$ to Dirac equation on the strip
 $0<y<\beta$
 \ben
 \left(\begin{array}{cr}
 \frac m2 & -\partial_z \\
 -\partial_{\bz} & \frac m2 \
 \end{array}\right)\left(\begin{array}{l}
 \psi_1  \\
 \psi_2 \
 \end{array}\right)=0,
 \ebn
 which has the following properties:
 \begin{itemize}
 \item The continuations of this solution to the left and right
   halfplane are quasiperiodic in $y$,
  \be\label{perl}
   \psi(x,y+\beta)=e^{2\pi i \lambda_0}\psi(x,y)\text{ for }x<0,
  \eb
  \be\label{perr}
   \psi(x,y+\beta)=e^{2\pi i \tilde{\lambda}\,\,\,}\psi(x,y)\text{ for }x>0,
  \eb
  where $\tilde{\lambda}=\lambda_0+\lambda_1$.
 \item It satisfies the normalization condition
 \be\label{norm}
 \lim_{|z|\rightarrow0}(mz/2)^{1-\lambda_1}\psi_1(x,y)=\frac{1}{\Gamma(\lambda_1)},
  \eb
 where the fractional power of $z$ is defined as
 \ben
 z^{1-\lambda_1}=e^{(1-\lambda_1)\ln z},\qquad 0< \mathrm{Im} (\ln z) <2\pi.
\ebn
 \end{itemize}
Theorem~\ref{teo1} shows that these requirements determine the
solution uniquely.

 The function $e^{-2\pi i\lambda_0 y/\beta}\psi$  is periodic in the left
 halfplane and therefore can be expanded there in Fourier
 series. Substituting the series into Dirac equation, one obtains a
 general form of the solution for $x<0$,
 \be\label{xm0}
 \psi_{x<0}(x,y)=-A\sum\limits_{n\in \Zb+\lambda_0}\frac{G(\theta_n)}{m\beta\cosh\theta_n}\,
 e^{mx\cosh\theta_n+imy\sinh\theta_n}
 \left(\begin{array}{c}
 e^{\theta_n} \\
 1
 \end{array}\right),
 \eb
 where
 $ \sinh\theta_n =\frac{2\pi}{m\beta}n $
 and the factor $1/(m\beta\cosh\theta_n)$ is introduced for further
 convenience. Analogously, the general form of solution in the right
 halfplane is
  \be\label{xb0}
 \psi_{x>0}(x,y)=A\sum\limits_{n\in \Zb-\tilde{\lambda}}\frac{H(\theta_n)}{m\beta\cosh\theta_n}\,
 e^{-mx\cosh\theta_n-imy\sinh\theta_n}
 \left(\begin{array}{c}
 -e^{\theta_n}\\
 1
 \end{array}\right).
 \eb
   \begin{figure}[h]
\begin{center}
 \includegraphics[height=4.5cm]{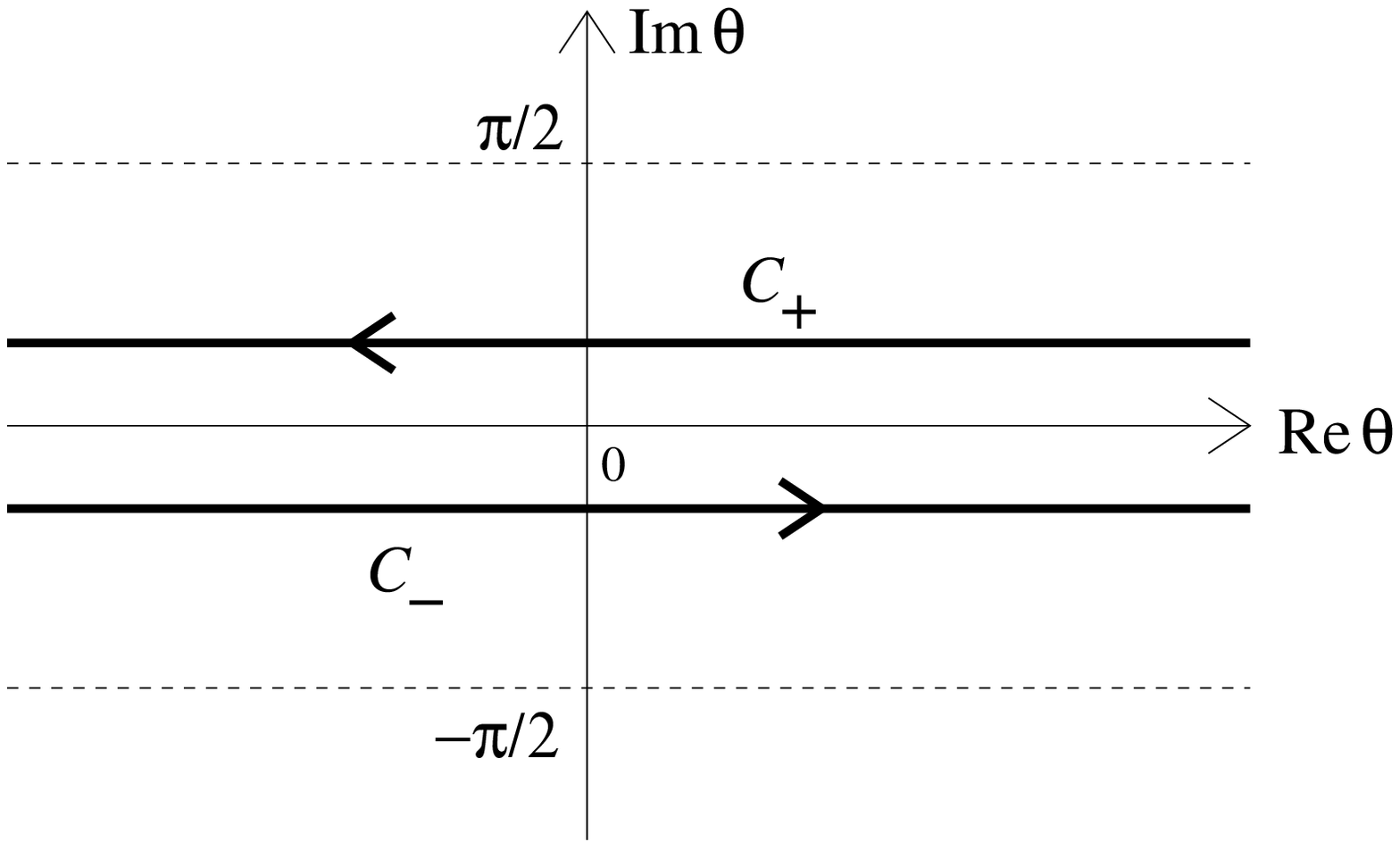}\vspace{0.5cm}

 Fig. 3. $\qquad$
 \end{center}
 \end{figure}

 Of course, in order for the series (\ref{xm0}) and (\ref{xb0}) to
 converge the functions $G(\theta)$ and $H(\theta)$ have not to
 grow too rapidly as $\theta\rightarrow \pm \infty$. Moreover, we shall
 assume that $G(\theta)$ and $H(\theta)$ are analytic in the strip
 $-\frac{\pi}{2}-\delta<\mathrm{Im}\,\theta< \frac{\pi}{2}+\delta$ for
 some $\delta>0$, so that (\ref{xm0}) and (\ref{xb0}) can be
 represented via contour integrals (see Fig.~3)
 \ben
  \psi_{x<0}(x,y)=A\int\nolimits_{C_- \bigcup C_+}\frac{d\theta}{2\pi}
  \frac{G(\theta)}{1-e^{im\beta\sinh\theta-2\pi i\lambda_0}}\,
  e^{mx\cosh\theta+imy\sinh\theta}\left(\begin{array}{c}
  e^{\theta} \\ 1
 \end{array}\right),
 \ebn
 \ben
  \psi_{x>0}(x,y)=A\int\nolimits_{C_- \bigcup C_+}\frac{d\theta}{2\pi}
  \frac{H(\theta)}{1-e^{-im\beta\sinh\theta-2\pi i\tilde{\lambda}\;}}\,
  e^{-mx\cosh\theta-imy\sinh\theta}\left(\begin{array}{c}
  -e^{\theta} \\ 1
 \end{array}\right).
 \ebn
 If $0<y<\beta$, the contours $C_+$ and $C_-$ can be continuously
 deformed into $\mathrm{Im}\,\theta=\frac{\pi}{2}$ and
 $\mathrm{Im}\,\theta=-\frac{\pi}{2}$, respectively, defining the
 continuations of $\psi_{x<0}(x,y)$ and $\psi_{x>0}(x,y)$ on the whole
 strip
 \begin{eqnarray}
 \nonumber\psi_{x<0}(x,y)=A\int\limits_{-\infty}^{\infty}\frac{d\theta}{2\pi}
 \left\{-\frac{G(\theta+i\pi/2)e^{imx\sinh\theta
       -my\cosh\theta}}{1-e^{-m\beta\cosh\theta-2\pi i\lambda_0}}
 \left(\begin{array}{c}
  ie^{\theta} \\ 1
 \end{array}\right)+\right.\\ \nonumber\left.+\frac{G(\theta-i\pi/2)e^{-imx\sinh\theta
       +my\cosh\theta}}{1-e^{m\beta\cosh\theta-2\pi i\lambda_0}}
 \left(\begin{array}{c}
  -ie^{\theta} \\ 1
 \end{array}\right)\right\},
 \end{eqnarray}
  \begin{eqnarray}
 \nonumber\psi_{x>0}(x,y)=A\int\limits_{-\infty}^{\infty}\frac{d\theta}{2\pi}
 \left\{-\frac{H(\theta+i\pi/2)e^{-imx\sinh\theta
       +my\cosh\theta}}{1-e^{m\beta\cosh\theta-2\pi i\tilde{\lambda}}}
 \left(\begin{array}{c}
  -ie^{\theta} \\ 1
 \end{array}\right)+\right.\\ \nonumber\left.+\frac{H(\theta-i\pi/2)e^{imx\sinh\theta
       -my\cosh\theta}}{1-e^{-m\beta\cosh\theta-2\pi i\tilde{\lambda}}}
 \left(\begin{array}{c}
 ie^{\theta} \\ 1
 \end{array}\right)\right\}.
 \end{eqnarray}
 These continuations coincide if two functional relations for
 $G(\theta)$ and $H(\theta)$ hold:
  \be\label{fr1}
  \frac{G(\theta+i\pi/2)}{H(\theta-i\pi/2)}=-\frac{1-e^{-m\beta\cosh\theta-2\pi i\lambda_0}}
  {1-e^{-m\beta\cosh\theta-2\pi i\tilde{\lambda}\;\;}}\,,
   \eb
  \be\label{fr2}
  \frac{G(\theta-i\pi/2)}{H(\theta+i\pi/2)}=-e^{2\pi i\lambda_1}\frac{1-e^{-m\beta\cosh\theta+2\pi i\lambda_0}}
  {1-e^{-m\beta\cosh\theta+2\pi i\tilde{\lambda}\;\;}}\, .
   \eb
 The relevant solutions of these equations
 can be found using the following lemma.
 \begin{lemma}\label{lem1} Consider two functions, $f(\theta)$ and
   $g(\theta)$, that are analytic in the strip
   $|\mathrm{Im}\,\theta|<\delta$. If in this strip
   $|f(\theta)|=O\left(\frac{1}{|\mathrm{Re}\,\theta|^2}\right)$ and $|g(\theta)|=O(1)$
 as $\mathrm{Re}\,\theta\rightarrow\pm\infty$, then the functions
 \be\label{etanu}
 \nu(\theta)=\int\limits_{-\infty}^{\infty}\frac{d\theta'}{2\pi}\,
 \tanh(\theta'-\theta)
 f(\theta'), \qquad
  \eta(\theta)=\int\limits_{-\infty}^{\infty}\frac{d\theta'}{2\pi}\,
 \mathrm{sech}\,(\theta'-\theta)
 g(\theta'), \qquad \theta\in\Rb
 \eb
 can be analytically continued to the strip
 $|\mathrm{Im}\,\theta|<\frac{\pi}{2}+\delta$. Furthermore, if
 $|\mathrm{Im}\,\theta|<\delta$, these continuations satisfy the
 relations
 \be\label{relations}
 \nu\Bigl(\theta+\frac{i\pi}{2}\Bigr)-\nu\Bigl(\theta-\frac{i\pi}{2}\Bigr)=-if(\theta),\qquad
 \eta\Bigl(\theta+\frac{i\pi}{2}\Bigr)+\eta\Bigl(\theta-\frac{i\pi}{2}\Bigr)=g(\theta).
 \eb
 \end{lemma}
 $\blacksquare$ Obviously, the expressions (\ref{etanu}) for
 $\nu(\theta)$ and $\eta(\theta)$ are
 analytic functions in the strip $|\mathrm{Im}\,\theta|<\frac{\pi}{2}$. Their
 analytic continuations to
 $|\mathrm{Im}\,\theta|<\frac{\pi}{2}+\delta$ are
 \ben
 \nu(\theta)\Bigl|_{\mathrm{Im}\,\theta=\pm\frac{\pi}{2}}=\mp\frac{i}{2}\,f(\theta\mp\frac{i\pi}{2})+
 P\!\,\int\limits_{-\infty}^{\infty}\frac{d\theta'}{2\pi}\,
 \coth\Bigl(\theta'-\theta\pm \frac{i\pi}{2}\Bigr) f(\theta'),
 \ebn
 \ben
 \eta(\theta)\Bigl|_{\mathrm{Im}\,\theta=\pm\frac{\pi}{2}}=\frac12\,g(\theta\mp\frac{i\pi}{2})
 \pm i\,P\!\!\int\limits_{-\infty}^{\infty}\frac{d\theta'}{2\pi}\,\,
 \mathrm{csch}\Bigl(\theta'-\theta\pm \frac{i\pi}{2}\Bigr) g(\theta'),
 \ebn
 \ben
 \nu(\theta)\Bigl|_{\frac{\pi}{2}<\mathrm{Im}\,\theta<\frac{\pi}{2}+\delta}=-if\Bigl(\theta-\frac{i\pi}{2}\Bigr)+
 \int\limits_{-\infty}^{\infty}\frac{d\theta'}{2\pi}\,
 \tanh(\theta'-\theta)
 f(\theta'),
 \ebn
 \ben
 \eta(\theta)\Bigl|_{\frac{\pi}{2}<\mathrm{Im}\,\theta<\frac{\pi}{2}+\delta}=g\Bigl(\theta-\frac{i\pi}{2}\Bigr)+
 \int\limits_{-\infty}^{\infty}\frac{d\theta'}{2\pi}\,
 \mathrm{sech}(\theta'-\theta)
 g(\theta'),
 \ebn
 \ben
 \nu(\theta)\Bigl|_{-\frac{\pi}{2}-\delta<\mathrm{Im}\,\theta<-\frac{\pi}{2}}=if\Bigl(\theta+\frac{i\pi}{2}\Bigr)+
 \int\limits_{-\infty}^{\infty}\frac{d\theta'}{2\pi}\,
 \tanh(\theta'-\theta)
 f(\theta'),
 \ebn
 \ben
 \eta(\theta)\Bigl|_{-\frac{\pi}{2}-\delta<\mathrm{Im}\,\theta<-\frac{\pi}{2}}=g\Bigl(\theta+\frac{i\pi}{2}\Bigr)+
 \int\limits_{-\infty}^{\infty}\frac{d\theta'}{2\pi}\,
 \mathrm{sech}(\theta'-\theta)
 g(\theta'),
 \ebn
 where the first two integrals are understood in the principal value
 sense. Then the statement of the lemma follows
 immediately. $\blacksquare$

 If we write the functions $G(\theta)$ and $H(\theta)$ in the form
 \be\label{auxi}
 \left\{\begin{array}{l}G(\theta)=-\exp\left(\pi
     i\lambda_1-\lambda_1\theta+\frac i2\,\nu(\theta)+\frac
   12\eta(\theta)\right),\\
 H(\theta)=\exp\left(-\lambda_1\theta+\frac i2\,\nu(\theta)-\frac
   12\eta(\theta)\right),\end{array}
 \right.
 \eb
 the functional relations (\ref{fr1}) and (\ref{fr2}) reduce to
 (\ref{relations}) with a particular choice of the functions
 $f(\theta)$ and $g(\theta)$:
 \be\label{rel1}
 f(\theta)=\ln
  \frac{1-e^{-m\beta\cosh\theta-2\pi i\lambda_0}}
  {1-e^{-m\beta\cosh\theta+2\pi i\lambda_0}}-\ln
  \frac{1-e^{-m\beta\cosh\theta-2\pi i\tilde{\lambda}}}
  {1-e^{-m\beta\cosh\theta+2\pi i\tilde{\lambda}}}\,,
 \eb
 \be\label{rel2}
 g(\theta)=\ln
  \frac{(1-e^{-m\beta\cosh\theta+2\pi i\lambda_0})(1-e^{-m\beta\cosh\theta-2\pi i\lambda_0})}
  {(1-e^{-m\beta\cosh\theta+2\pi i\tilde{\lambda}\;\;})(1-e^{-m\beta\cosh\theta-2\pi i\tilde{\lambda}\;\;})}.
 \eb
 The branches of logarithms in (\ref{rel1}) and  (\ref{rel2}) are fixed so that for real
 $\theta$ their
 imaginary parts lie in the interval $(-\pi;\pi)$.

 The formulae
 (\ref{xm0}), (\ref{xb0}), (\ref{etanu}) and
 (\ref{auxi})--(\ref{rel2}) provide a solution to the Dirac equation on
 the 1-punctured cylinder with specified branching (one should go back
 and check that all formal manipulations we have made with functions
 $G(\theta)$ and $H(\theta)$ indeed can be done). It remains only to
 verify the normalization condition (\ref{norm}).

 Let us take, say, the expansion (\ref{xb0}) and rewrite it using
 the
 Poisson summation formula:
 \ben
 \psi_{x>0}(x,y)=A\sum\limits_{k\in\Zb}\;\int\limits_{-\infty}^{\infty}\frac{d\theta}{2\pi}\,H(\theta)\,e^{
  -mx\cosh\theta-im(y+k\beta)\sinh\theta-2\pi i k\tilde{\lambda} }\left(\begin{array}{c}
  -e^{\theta} \\ 1
 \end{array}\right).
 \ebn
 The asymptotics of $\psi$ for $|z|\rightarrow0$ is determined by the term with $k=0$. Since
 $\lambda_1>0$, the main contribution to the corresponding integral is
 due to large $|\theta|$. Straightforward calculation shows that for $|z|\rightarrow0$
 \ben
 \psi_1(x,y)\sim -\frac{A}{2\pi}\,e^{-\pi i \lambda_1/2+
   i\nu_{\infty}/2}\Gamma (1-\lambda_1)(mz/2)^{\lambda_1-1},
 \ebn
 where
 \be\label{nuinf}
 \nu_{\infty}=\lim_{\theta\rightarrow+\infty}\nu(\theta)=-\frac{1}{2\pi}\int\limits_{-\infty}^{\infty}
 d\theta\left(\ln
  \frac{1-e^{-m\beta\cosh\theta-2\pi i\lambda_0}}
  {1-e^{-m\beta\cosh\theta+2\pi i\lambda_0}}-\ln
  \frac{1-e^{-m\beta\cosh\theta-2\pi i\tilde{\lambda}}}
  {1-e^{-m\beta\cosh\theta+2\pi i\tilde{\lambda}}} \right).
 \eb
 Therefore, the solution we have constructed differs from the element
 of the canonical basis only by a constant factor which can be set to unity
 by appropriate choice of $A$. Summarizing all these results,
 we obtain
 \begin{thm}\label{teo2}
 The element of canonical basis on the cylinder with one branchpoint
 is given by the following expressions:
 \ben
 \mathbf{w}(x,y)=A\sum\limits_{n\in\Zb+\lambda_0}\frac{e^{\pi i
       \lambda_1+\frac
     i2\nu(\theta_n)+\frac12\eta(\theta_n)}}{m\beta\cosh\theta_n}\,
 e^{-\lambda_1\theta_n+mx\cosh\theta_n+imy\sinh\theta_n}\left(\begin{array}{c}
 e^{\theta_n} \\ 1 \end{array}\right)\text{ for } x<0,
 \ebn
 \ben
 \mathbf{w}(x,y)=A\sum\limits_{n\in\Zb-\tilde{\lambda}}\frac{e^{\frac
     i2\nu(\theta_n)-\frac12\eta(\theta_n)}}{m\beta\cosh\theta_n}\,
 e^{-\lambda_1\theta_n-mx\cosh\theta_n-imy\sinh\theta_n}\left(\begin{array}{c}
 -e^{\theta_n} \\ 1 \end{array}\right)\text{ for } x>0,
 \ebn
 where the functions $\nu(\theta)$, $\eta(\theta)$ are determined from
 \ben
 \nu(\theta)=\int\limits_{-\infty}^{\infty}\frac{d\theta'}{2\pi} \,\tanh(\theta'-\theta)
  \left(\ln
  \frac{1-e^{-m\beta\cosh\theta'-2\pi i\lambda_0}}
  {1-e^{-m\beta\cosh\theta'+2\pi i\lambda_0}}-\ln
  \frac{1-e^{-m\beta\cosh\theta'-2\pi i\tilde{\lambda}}}
  {1-e^{-m\beta\cosh\theta'+2\pi i\tilde{\lambda}}} \right),
 \ebn
 \ben
 \eta(\theta)=\int\limits_{-\infty}^{\infty}\frac{d\theta'}{2\pi}
 \,\mathrm{sech}\,(\theta'-\theta)
 \ln  \frac{(1-e^{-m\beta\cosh\theta'+2\pi i\lambda_0})(1-e^{-m\beta\cosh\theta'-2\pi i\lambda_0})}
  {(1-e^{-m\beta\cosh\theta'+2\pi i\tilde{\lambda}\;\;})(1-e^{-m\beta\cosh\theta'-2\pi i\tilde{\lambda}\;\;})},
 \ebn
 and normalization constant
 $ A=-2\sin\pi\lambda_1 \,e^{-i\nu_{\infty}/2}$.
 \end{thm}
 \section{Green function for the Dirac operator}
 \subsection{Green function for $n=0$}
 Let us calculate the Green function on the
 cylinder without branchpoints. The domain of the Dirac operator in
 this case consists of quasiperiodic functions
 $\psi(x,y+\beta)=e^{2\pi i \lambda_0}\psi(x,y)$ that
 are square integrable in the strip $S=\{(x,y):0<y<\beta\}$. After
 Fourier transformation
 \ben
 \hat{\psi}{}(\xi_x,\xi_y)=\frac{1}{(2\pi)^2}\int\limits_{-\infty}^{\infty}dx\int\limits_0^{\beta}dy
 \;\psi(x,y)\,e^{-i(x\xi_x+y\xi_y)},\qquad \xi_x\in\Rb,\;\xi_y\in\frac{2\pi}{\beta}(\Zb+\lambda_0)
 \ebn
 the Dirac operator and its inverse are represented by matrices
 \ben
 D=\frac12\left(\begin{array}{cc}
 m & -i\bar{\xi} \\ -i\xi & m
\end{array}\right),\qquad
 D^{-1}=\frac{2}{m^2+|\xi|^2}\left(\begin{array}{cc}
 m & i\bar{\xi} \\ i\xi & m \end{array}\right),
 \ebn
 where $\xi=\xi_x+i\xi_y$, $\bar{\xi}=\xi_x-\xi_y$. Since the inverse
 transformation is given by
 \ben
 \psi(x,y)=\frac{2\pi}{\beta}\sum\limits_{\xi_y}\int\limits_{-\infty}^{\infty}d\xi_x \;\hat{\psi}{}(\xi_x,\xi_y)\,e^{i(x\xi_x+y\xi_y)},
 \ebn
 one obtains the following formula for the Green function
 $G_0(x-x',y-y')$ :
 \be\label{green00}
 G_0(x,y)=\frac{1}{\pi\beta}\sum\limits_{\xi_y}\int\limits_{-\infty}^{\infty}d\xi_x\left(\begin{array}{cc}
 m & i\bar{\xi} \\ i\xi & m \end{array}\right)\frac{e^{i(x\xi_x+y\xi_y)}}{m^2+|\xi|^2}.
 \eb
 Another two representations of the Green function will be useful for
 us.  Choose, for example, $x>0$ and calculate the integrals in
 (\ref{green00}):
 \be\label{green000}
 G_0(x,y)=\sum\limits_{n\in\Zb-\lambda_0}\frac{e^{-mx\cosh\theta_n-imy\sinh\theta_n}}{\beta\cosh\theta_n}\left(
 \begin{array}{cc}
 1 & -e^{\theta_n} \\
 -e^{-\theta_{n}} & 1
 \end{array}\right)=
 \eb
 \be\label{green0000} =m\int\nolimits_{C_-\cup
 C_+}\frac{d\theta}{2\pi}\frac{e^{-mx\cosh\theta-imy\sinh\theta}}{1-e^{-im\beta\sinh\theta-2\pi i\lambda_0}}\left(
 \begin{array}{cc}
 1 & -e^{\theta} \\
 -e^{-\theta} & 1
 \end{array}\right).
 \eb
 Analogously, for $x<0$ one obtains
 \be\label{green001}
 G_0(x,y)=\sum\limits_{n\in\Zb+\lambda_0}\frac{e^{mx\cosh\theta_n+imy\sinh\theta_n}}{\beta\cosh\theta_n}\left(
 \begin{array}{cc}
 1 & e^{\theta_n} \\
 e^{-\theta_{n}} & 1
 \end{array}\right)=
 \eb
 \be\label{green0001} =-m\int\nolimits_{C_-\cup
 C_+}\frac{d\theta}{2\pi}\frac{e^{mx\cosh\theta +imy\sinh\theta}}{1-e^{im\beta\sinh\theta-2\pi i\lambda_0}}\left(
 \begin{array}{cc}
 1 & e^{\theta} \\
 e^{-\theta} & 1
 \end{array}\right).
 \eb

 In what follows, we shall also need the asymptotics of these expressions as
 $x,y\rightarrow0$. To find it, let us rewrite (\ref{green00}) using
 the Poisson formula:
 \be\label{poisson1}
 G_0(x,y)=\frac{1}{2\pi^2}\sum\limits_{k\in\Zb}\;\int\limits_{-\infty}^{\infty}\int\limits_{-\infty}^{\infty}d\xi_xd\xi_y
 \left(\begin{array}{cc}
 m & i\bar{\xi} \\ i\xi & m \end{array}\right)\frac{e^{i(x\xi_x+y\xi_y)+ik(\beta\xi_y-2\pi\lambda_0)}}{m^2+|\xi|^2}.
 \eb
 The leading term of asymptotics is determined by the integral
 corresponding to $k=0$, representing the Green function on the
 plane. A little calculation shows that as $|z|\rightarrow0$
 \ben
 G_0(x,y) \sim -\frac{m}{\pi}\left(\begin{array}{cc}
 \ln |z| & 1/z \\
 1/\bz & \ln|z|
\end{array}\right).
 \ebn
 It is very convenient to write Green function as
 \ben
 G_0(z)=2G(z)J,\qquad J=\left(\begin{array}{cc}
 0 & i \\ -i & 0 \end{array}\right),
 \ebn
 since the rows of $G(z)$ satisfy the Dirac equation (and not its adjoint,
 as the rows of $G_0(z)$ do).

 \subsection{General properties of Green function}
 The domain $\mathcal{D}^{a,\lambda}$ of the Dirac operator
 $D^{a,\lambda}$ is chosen to consist of
 functions $\psi$ that have monodromies $e^{2\pi i \lambda_{\nu}}$
 ($\nu=0,\ldots,n$)
 and are integrable at $|x|\rightarrow\infty$. We also
 require (see (\ref{bcond}))
 \be\label{bcond2}
 \left(\begin{array}{cc}(z-a_{\nu})^{-\lambda_{\nu}}&0\\
 0&(\bz-\bar{a}_{\nu})^{\lambda_{\nu}}\end{array}\right)\psi[a_{\nu}]\in H^1[a_{\nu}], \qquad \nu=1,\ldots,n
 \eb
 where $H^1[a_{\nu}]$ denotes the space of functions that are square
 integrable in the neighborhood of $\{a_{\nu}\}$ together with
 their first derivatives. In the previous
 section we have shown that the Dirac equation $D^{a,\lambda}\psi=0$ has no
 solutions in $\mathcal{D}^{a,\lambda}$. Thus ``naively'' we can
 think of $D^{a,\lambda}$ as being an invertible operator. The kernel of the
 inverse is called the
 Green function $G^{a,\lambda}$. More precisely, the solution of
 \ben
 D^{a,\lambda}\psi =\varphi,
 \ebn
 is assumed to have the form
 \be\label{green}
 \psi(z)=\int\limits_{\Cy\backslash
 b}G^{a,\lambda}(z,z')J\,\varphi(z')\,idz'\wedge d\bar{z'},
 \eb
 Then one can try to determine the Green function by the following requirements:
 \begin{itemize}
 \item The columns of $G^{a,\lambda}(z,z')$ must satisfy Dirac equation
 $D_{z}G_{\cdot,j}^{a,\lambda}(z,z')=0$ for all
 $z\in\Cy\backslash( b\cup\{z'\})$; they are square integrable functions at $|x|\rightarrow\infty$
 that have monodromy $e^{2\pi i \lambda_{\nu}}$ ($\nu=0,\ldots,n$) and singular behaviour
  (\ref{bcond2}) at each singularity. Therefore,
  \be\label{green1}
 G^{a,\lambda}_{\cdot,j}(z,z')[a_{\nu}]=\sum\limits_{k>0}\left\{
 a^{\;(\nu)}_{k,j}(z')w_{k+\lambda_{\nu}}[a_{\nu}]+
 b^{\;(\nu)}_{k,j}(z')w_{k-\lambda_{\nu}}^{\;*}[a_{\nu}]\right\}.
  \eb
  \item The integral operator with the kernel
  $D_z^{a,\lambda}G^{a,\lambda}(z,z')$ has to ``cut out'' the values of the function
  $\varphi(z)$. Therefore, the singular behaviour of $G^{a,\lambda}(z,z')$
  for $z\rightarrow z'$ must coincide with that of Green function
  for the Dirac operator on the cylinder without branchpoints,
  \be\label{green0}
  G^{a,\lambda}(z,z')-G(z,z')\in C^1(z\rightarrow z').
  \eb
 \end{itemize}
 \textbf{Remark}. Suppose the function defined by these conditions exists. Then it is
 unique, since the columns of the difference of two such functions are obviously
 in $\widetilde{\mathbf{W}}^{a,\lambda}$. \vspace{0.3cm}

 Next we determine how $G^{a,\lambda}(z, z')$ depends on the second argument.
 To do this, let us define a matrix $F^{a,\lambda}(z,z')$
 satisfying the following conditions:
 \begin{itemize}
 \item The rows of $F^{a,\lambda}(z,z')$ satisfy Dirac equation
 $D_{z'}G_{j,\cdot}^{a,\lambda}(z,z')=0$ for all
 $z'\in\Cy\backslash( b\cup\{z\})$; they are square integrable functions at $|x|\rightarrow\infty$,
 have the inverse monodromy $e^{-2\pi i \lambda_{\nu}}$
 ($\nu=0,\ldots,n$) and corresponding
  singular behaviour (\ref{bcond2}) at the branching points:
   \be\label{green2}
 F^{a,\lambda}_{j,\cdot}(z,z')[a_{\nu}]=\sum\limits_{k>0}\left\{
 \alpha^{\;(\nu)}_{k,j}(z)w_{k-\lambda_{\nu}}[a_{\nu}]+
 \beta^{\;(\nu)}_{k,j}(z)w_{k+\lambda_{\nu}}^{\;*}[a_{\nu}]\right\}.
  \eb
  \item The singular behaviour of $F^{a,\lambda}(z,z')$
  for $z'\rightarrow z$ coincides with the asymptotics of ``unperturbed'' Green function,
  \be\label{green3}
  F^{a,\lambda}(z,z')-G(z,z')\in C^1(z'\rightarrow z).
  \eb
 \end{itemize}
 Skipping the proof of existence of $G^{a,\lambda}(z,z')$ (or
 $F^{a,\lambda}(z,z')$) we shall now prove the following
 \begin{thm}
 $G^{a,\lambda}(z,z')=F^{a,\lambda}(z,z')$.
 \end{thm}
 $\blacksquare$  First we note an auxiliary relation: if $f(z)$ and $g(z)$ are
 smooth functions on some open set $U\subset\Cy$, then
 \be\label{stokes2}
 \left\{Df\cdot Jg-f\cdot JDg\right\}dz\wedge d\bz=\bigl\{
 \partial_z(f_2g_2)-\partial_{\bz}(f_1g_1)\bigr\}\,dz\wedge d\bz=
 d(f_1 g_1 dz+ f_2 g_2 d\bz).
 \eb
 Now choose two distinct points $x,y\notin b$ and
 \ben
 \begin{cases}
 f(z)=F^{a,\lambda}_{i,\cdot}(x,z), \\
 g(z)=G^{a,\lambda}_{\cdot,j}(z,y), \
  \end{cases}
 \ebn
 with $U$ being the complement to the union of the disks $\left(\bigcup\limits_{\nu=1}^n
 D_{\varepsilon}(a_{\nu})\right)\cup D_{\varepsilon}(x)\cup
 D_{\varepsilon}(y)$ and two strips: $S'_{\varepsilon}=\{(x,y):0\leq
 y<\varepsilon\}$ and
 $S''_{\varepsilon}=\{(x,y):\beta-\varepsilon<y
 \leq\beta\}$. Integrating (\ref{stokes2}) over this set and
 using Stokes theorem, one obtains
 \ben
 0=\sum\limits_{\nu=1}^n
 \oint\limits_{ \partial D_{\varepsilon}(a_{\nu})}\left(f_1g_1dz+f_2g_2d\bz\right)+\!\!\!\!\!\!\!\!
 \oint\limits_{ \partial D_{\varepsilon}(x)\cup\partial D_{\varepsilon}(y)}\!\!\!\! \!\!\!\!\left(f_1g_1dz+f_2g_2d\bz\right)+
\!\!\!\! \oint\limits_{ \partial S_{\varepsilon}'\cup\partial S_{\varepsilon}''}
  \!\!\!\!\left(f_1g_1dz+f_2g_2d\bz\right).
 \ebn
 The expression under the integrals is single-valued on $\Cy\backslash
 a$, so in the limit $\varepsilon\rightarrow0$ the last integral
 cancels out.  The integrals over $ \partial D_{\varepsilon}(a_{\nu})$
 also vanish. One can easily check this, substituting instead of
 $G^{a,\lambda}_{\cdot,j}(z,y)$ and $F^{a,\lambda}_{i,\cdot}(x,z)$
 their local expansions
 (\ref{green1}) and (\ref{green2}). It remains to calculate
 \ben
 \lim_{\varepsilon\rightarrow0}\biggl\{\oint\limits_{ \partial D_{\varepsilon}(x)}
 F^{a,\lambda}_{i,1}(x,z)G^{a,\lambda}_{1,j}(z,y)dz+
 \oint\limits_{ \partial D_{\varepsilon}(x)}
 F^{a,\lambda}_{i,2}(x,z)G^{a,\lambda}_{2,j}(z,y)d\bz\biggr\}=
 \ebn
 \ben
 =\frac{i}{2\pi}\left\{-2\pi i\delta_{i1}G^{a,\lambda}_{1,j}(x,y)-
 2\pi
 i\delta_{i2}G^{a,\lambda}_{2,j}(x,y)\right\}=G^{a,\lambda}_{i,j}(x,y),
 \ebn
 and, similarly,
 \ben
 \lim_{\varepsilon\rightarrow0}\biggl\{\oint\limits_{ \partial D_{\varepsilon}(y)}
 F^{a,\lambda}_{i,1}(x,z)G^{a,\lambda}_{1,j}(z,y)dz+
 \oint\limits_{ \partial D_{\varepsilon}(y)}
 F^{a,\lambda}_{i,2}(x,z)G^{a,\lambda}_{2,j}(z,y)d\bz\biggr\}=-F^{a,\lambda}_{i,j}(x,y).
 \ebn
 Finally we get
 $G^{a,\lambda}_{i,j}(x,y)=F^{a,\lambda}_{i,j}(x,y)$.
 $\blacksquare$\vspace{0.3cm}\\
 \textbf{Remark}. Since the columns and rows of
 $G^{a,\lambda}(z,z')$ satisfy Dirac equation in $z$ and $z'$
 respectively, so do also the derivatives of the Green function
 $\partial_{a_{\mu}}G^{a,\lambda}(z,z')$ and
 $\partial_{\bar{a}_{\mu}}G^{a,\lambda}(z,z')$. These derivatives
 are not singular at $z\rightarrow z'$. However, their local
 expansions are more singular than (\ref{green1}) or
 (\ref{green2}); from the relations
 \ben
 \begin{cases}
 \partial_z w_l=\frac m2 w_{l-1}, \\
 \partial_{\bz} w_l=\frac m2 w_{l+1},
 \end{cases}\qquad
  \begin{cases}
 \partial_z w^*_l=\frac m2 w^*_{l+1}, \\
 \partial_{\bz} w^*_l=\frac m2 w^*_{l-1}
 \end{cases}
 \ebn
 it follows that
 \be\label{ll4}
 \partial_{a_{\mu}}G_{j,\cdot}^{a,\lambda}(z,z')[a_{\nu}]=-\frac
 m2\,\delta_{\mu\nu}\alpha_{1/2,j}^{\;(\nu)}(z)w_{-1/2-\lambda_{\nu}}+
 \sum\limits_{k>0}\left\{\gamma_{k,j}^{\;(\nu)}(z)w_{k-\lambda_{\nu}}+
 \tilde{\gamma}_{k,j}^{\;(\nu)}(z)w^{\;*}_{k+\lambda_{\nu}}\right\},
 \eb
  \be\label{ll5}
 \partial_{\bar{a}_{\mu}}G_{j,\cdot}^{a,\lambda}(z,z')[a_{\nu}]=-\frac
 m2\,\delta_{\mu\nu}\beta_{1/2,j}^{\;(\nu)}(z)w^{\;*}_{-1/2+\lambda_{\nu}}+
 \sum\limits_{k>0}\left\{\eta_{k,j}^{\;(\nu)}(z)w_{k-\lambda_{\nu}}+
 \tilde{\eta}_{k,j}^{\;(\nu)}(z)w^{\;*}_{k+\lambda_{\nu}}\right\}.
 \eb
 Therefore, if we somehow determine the coefficients
 $\alpha_{1/2,j}^{\;(\nu)}(z)$, $\beta_{1/2,j}^{\;(\nu)}(z)$ and
 find the solution of Dirac equation with the same ``extra''
 singular behaviour, it will coincide with the corresponding
 derivative of Green function.\vspace{0.3cm}

 Let us consider a multivalued solution $f(z)$ to the Dirac equation in the strip
 $S=\{(x,y):0<y<\beta\}$, which is square
 integrable at $|x|\rightarrow\infty$ and has the following local expansions about the
 branchpoints $a_{\nu}$ (there are no other singularities!)
 \ben
 f(z')[a_{\nu}]=\sum_{k}\left\{a_k^{(\nu)} w_{k-\lambda_{\nu}}+
 b_k^{(\nu)} w_{k+\lambda_{\nu}}^{\;*}\right\}.
 \ebn
 In addition, only a finite number of coefficients
 $a^{(\nu)}_k$, $b^{(\nu)}_k$ with
 $k<0$ is allowed to have non-zero values. Lower and upper
 continuations of $f(z)$ across the branchcut $d_{\nu}$ ($\nu=0,\ldots,n$) are supposed to
 differ by monodromy multiplier $\exp\bigl({2\pi i \sum\limits_{k=0}^{\nu}\lambda_{k}}\bigr)$, so that the product
 $G_{j,\cdot}^{a,\lambda}(z,z')\overline{f(z')}$, as the function of
 $z'$, is  single-valued
  on $\Cy\backslash a$. Using (\ref{stokes1}) and Stokes
 theorem, one obtains
 \be\label{ll0}
 \frac m2\int\limits_{\Cy\backslash\bigcup\limits_{\nu}D_{\varepsilon}(a_{\nu})}
 G_{j,\cdot}^{a,\lambda}(z,z')\overline{f(z')} i dz'\wedge
 d\bar{z'}=
 \eb
 \be\label{ll1}
 =-\sum\limits_{\nu}\oint\limits_{\partial D_{\varepsilon}(a_{\nu})}
 G_{j,2}^{a,\lambda}(z,z')\overline{f_1(z')}id\bar{z'}-
 \oint\limits_{\partial D_{\varepsilon}(z)}
 G_{j,2}^{a,\lambda}(z,z')\overline{f_1(z')}id\bar{z'}=
 \eb
  \be\label{ll2}
 =\sum\limits_{\nu}\oint\limits_{\partial D_{\varepsilon}(a_{\nu})}
 G_{j,1}^{a,\lambda}(z,z')\overline{f_2(z')}idz'+
 \oint\limits_{\partial D_{\varepsilon}(z)}
 G_{j,1}^{a,\lambda}(z,z')\overline{f_2(z')}idz'.
 \eb
 The surface integral (\ref{ll0}) does not converge for
 $\varepsilon\rightarrow0$. However, one can still compare the
 asymptotics of (\ref{ll1}) and (\ref{ll2}) for
 $\varepsilon\rightarrow0$:
 \ben
 \lim_{\varepsilon\rightarrow0}\text{(\ref{ll1})}=-\frac 4m
 \sum\limits_{\nu}\sum\limits_{k>0}(-1)^{k+1/2}\sin\pi\lambda_{\nu}\,
 \beta_{k,j}^{\;(\nu)}(z)\overline{a_{-k}^{(\nu)}}
 -i\delta_{j2}\overline{f_1(z)},
 \ebn
  \ben
 \lim_{\varepsilon\rightarrow0}\text{(\ref{ll2})}=\frac 4m
 \sum\limits_{\nu}\sum\limits_{k>0}(-1)^{k+1/2}\sin\pi\lambda_{\nu}\,
 \alpha_{k,j}^{\;(\nu)}(z)\overline{b_{-k}^{(\nu)}}
 +i\delta_{j1}\overline{f_2(z)}.
 \ebn
  Finally we get
 \be\label{ll3}
 \sum\limits_{\nu}\sum\limits_{k>0}(-1)^{k+1/2}\sin\pi\lambda_{\nu}\,\left\{
 \beta_{k,j}^{\;(\nu)}(z)\overline{a_{-k}^{(\nu)}}
 +\alpha_{k,j}^{\;(\nu)}(z)\overline{b_{-k}^{(\nu)}}\right\}=
 -\frac{im}{4}\bigl(\delta_{j2}\overline{f_1(z)}+\delta_{j1}\overline{f_2(z)}\,\bigr).
 \eb
 Now we shall make a special choice of $f(z')$ to find the lowest
 coefficients of the Green function expansions. Analogously to the
 previous section (see (\ref{basis})), let us introduce $n$ special multivalued solutions
 to Dirac equation, that are integrable at $|x|\rightarrow\infty$ and have local
 expansions
 \be\label{basis2}
 \widetilde{\mathbf{w}}_{\mu}(\lambda)[a_{\nu}]=
 \delta_{\mu\nu}w_{-1/2+\lambda_{\nu}}[a_{\nu}]+
 \sum\limits_{k>0}\left\{a_{\;k}^{(\nu)}
 (\widetilde{\mathbf{w}}_{\mu}(\lambda)) w_{k+\lambda_{\nu}}[a_{\nu}]+
 b_{\;k}^{(\nu)}(\widetilde{\mathbf{w}}_{\mu}(\lambda))
  w_{k-\lambda_{\nu}}^{\;*}[a_{\nu}]\right\}
 \eb
 The existence and uniqueness of such solutions follow from the
 existence of canonical basis; two bases coincide if all
 $\lambda_{\nu}>0$.

 After the substitution $f(z)=\widetilde{\mathbf{w}}_{\mu}(-\lambda)$ in
 (\ref{ll3}) one obtains
 \be\label{alpha12}
 \left(\begin{array}{l}
 \beta_{1/2,1}^{\;(\mu)}(z) \\
  \beta_{1/2,2}^{\;(\mu)}(z)
 \end{array}\right)=\frac{im}{4\sin\pi\lambda_{\mu}}
 \left(\begin{array}{l}
 \overline{\widetilde{\mathbf{w}}_{\mu2}(z,-\lambda)} \\
 \overline{\widetilde{\mathbf{w}}_{\mu1}(z,-\lambda)}
 \end{array}\right)=\frac{im}{4\sin\pi\lambda_{\mu}}\,
 \widetilde{\mathbf{w}}_{\mu}^*(z,-\lambda).
 \eb
 On the other hand, the substitution $f(z)=
 \widetilde{\mathbf{w}}_{\mu}^*(z,\lambda)$ in (\ref{ll3}) gives
 a formula for $\alpha_{1/2,j}^{\;(\mu)}(z)$,
  \be\label{beta12}
 \left(\begin{array}{l}
 \alpha_{1/2,1}^{\;(\mu)}(z) \\
  \alpha_{1/2,2}^{\;(\mu)}(z)
 \end{array}\right)=\frac{im}{4\sin\pi\lambda_{\mu}}\,
 \widetilde{\mathbf{w}}_{\mu}(z,\lambda).
 \eb
 Taking into  account the earlier remarks (the
 formulae (\ref{ll4}) and (\ref{ll5})), we obtain an expression
 for the derivatives of the Green function in terms of the solutions
 (\ref{basis2}):
 \be\label{der1}
 \partial_{a_j}G^{a,\lambda}(z,z')=-\frac{im^2}{8\sin\pi\lambda_j}\,
 \widetilde{\mathbf{w}}_{j}(z,\lambda)\otimes\widetilde{\mathbf{w}}_{j}(z',-\lambda),
 \eb
 \be\label{der2}
 \partial_{\bar{a}_j}G^{a,\lambda}(z,z')=-\frac{im^2}{8\sin\pi\lambda_j}\,
 \widetilde{\mathbf{w}}_{j}^*(z,-\lambda)\otimes\widetilde{\mathbf{w}}_{j}^*(z',\lambda).
 \eb
 \subsection{One-point Green function}
 The formulae (\ref{der1}) and (\ref{der2}) can be used to calculate
 the Green function $G^{a,\lambda}$ on the cylinder with one branching point
 $\{a\}$. Notice that in the case of a single puncture  the
 solutions  (\ref{basis2}) are easily expressed via the element of
 canonical basis. If we suppose for definiteness that
 $0<\lambda_1<\frac12$ then
 $\tilde{\mathbf{w}}(z,\lambda)=\mathbf{w}(z,\lambda)$. Differentiating the local expansions, one can also verify that
 \ben
 \tilde{\mathbf{w}}(z,-\lambda)=\frac{2}{m}\partial_{z}\mathbf{w}(z,1-\lambda).
 \ebn
 Let us determine  $G^{a,\lambda}(z,z')$ if both $z,z'\in\Cy\backslash
 b$ are in the left
 half-strip: $\mathrm{Re}\,z,\mathrm{Re}\,z'<\mathrm{Re}\,a$. Using the results of the
 previous section, one can write
 \be\label{eq31}
 \tilde{\mathbf{w}}(z,\lambda)=-A(\lambda)\sum\limits_{l\in\Zb+\lambda_0}
\frac{G(\theta_l;\lambda)e^{\frac m2 (z-a)e^{\theta_l}+\frac m2 (\bz-\bar{a})e^{-\theta_l}}}{m\beta\cosh\theta_l}
  \left(\begin{array}{l}
 e^{\theta_l} \\
   1 \end{array}\right),\eb
  \be\label{eq32}
 \tilde{\mathbf{w}}(z,-\lambda)=-A(1-\lambda)\sum\limits_{n\in\Zb-\lambda_0}
\frac{G(\theta_n;1-\lambda)e^{\theta_n+\frac m2 (z-a)e^{\theta_n}+\frac m2 (\bz-\bar{a})e^{-\theta_n}}}{m\beta\cosh\theta_n}
  \left(\begin{array}{l}
 e^{\theta_n} \\
   1 \end{array}\right).\eb
 Note the simple relations
 \ben
 A(\lambda)A(1-\lambda)=4\sin^2\pi\lambda_1, \ebn \be\label{rel0}
 \nu(\theta;\lambda)=-\nu(\theta;1-\lambda)=-\nu(\theta;-\lambda),
 \eb \ben
 \eta(\theta;\lambda)=\eta(\theta;1-\lambda)=\eta(\theta;-\lambda).
 \ebn
 If we substitute (\ref{eq31}) and (\ref{eq32}) into the equations
 (\ref{der1}),  (\ref{der2}) and then integrate them, we obtain the
 following formula for Green function $G^{a,\lambda}(z,z')$:
  \be\label{eq33}
 G^{a,\lambda}(z,z')=i\sin\pi\lambda_1\sum\limits_{l\in\Zb+\lambda_0}\sum\limits_{n\in\Zb-\lambda_0}
 \frac{G(\theta_l;\lambda)G(\theta_n;1-\lambda)e^{\theta_n}}
  {m\beta^2\cosh\theta_l\cosh\theta_n}\times
\eb
 \ben
 \times \frac{e^{\frac m2\{ (z-a)e^{\theta_l}+
   (\bz-\bar{a})e^{-\theta_l}+ (z'-a)e^{\theta_n}+
 (\bar{z}'-\bar{a})e^{-\theta_n}\}}}{e^{\theta_l}+e^{\theta_n}} \left(\begin{array}{cc}
 e^{\theta_l+\theta_n} & e^{\theta_l} \\
  e^{\theta_n} & 1\end{array}\right)+ C(z,z').
\ebn
 The function $C(z,z')$ does not depend on the position of branching
 point and has to be determined. We remark that the double sum
 in the RHS of the last relation converges even if $z=
 z'$. This gives a hint that  $C(z,z')$ should be equal to
 ``unperturbed'' Green function $G(z-z';\lambda_0)$. Assuming
 this is indeed the case, let us  fix $a=0$ and rewrite
 (\ref{eq33})  via contour integrals
  \ben
 G^{0,\lambda}(z,z')=
 im\sin\pi\lambda_1\!\!\!\!\int\limits_{C_-\cup
   C_+}\!\!\!\!\frac{d\theta}{2\pi}\int\limits_{C_-\cup   C_+}\!\!\!\!
 \frac{d\theta'}{2\pi}\;\frac{G(\theta;\lambda)G(\theta';1-\lambda)e^{\theta'}}
  {(1-e^{im\beta\sinh\theta-2\pi i
      \lambda_0})(1-e^{im\beta\sinh\theta'+2\pi i
      \lambda_0})} \times
 \ebn
 \be\label{eq34}
 \times
\frac{e^{\frac m2\{ ze^{\theta}+
   \bz e^{-\theta}+ z'e^{\theta'}+
 \bar{z}'e^{-\theta'}\}}}{e^{\theta}+e^{\theta'}}
 \left(\begin{array}{cc}
 e^{\theta+\theta'} & e^{\theta} \\
  e^{\theta'} & 1\end{array}\right)+G(z-z';\lambda_0).
\eb
Analogously, if $\mathrm{Re}\,z, \mathrm{Re}\,z'>0$, the Green function
$G^{0,\lambda}(z,z')$ is assumed to have the following form:
  \ben
 G^{0,\lambda}(z,z')=
 im\sin\pi\lambda_1\!\!\!\!\int\limits_{C_-\cup
   C_+}\!\!\!\!\frac{d\theta}{2\pi}\int\limits_{C_-\cup   C_+}\!\!\!\!
 \frac{d\theta'}{2\pi}\;\frac{H(\theta;\lambda)H(\theta';1-\lambda)e^{\theta'}}
  {(1-e^{-im\beta\sinh\theta-2\pi i
      \tilde{\lambda}})(1-e^{-im\beta\sinh\theta'+2\pi i
      \tilde{\lambda}})} \times
 \ebn
 \be\label{eq35}
 \times
\frac{e^{-\frac m2\{ ze^{\theta}+
   \bz e^{-\theta}+ z'e^{\theta'}+
 \bar{z}'e^{-\theta'}\}}}{e^{\theta}+e^{\theta'}}
 \left(\begin{array}{cc}
 e^{\theta+\theta'} & -e^{\theta} \\
 - e^{\theta'} & 1\end{array}\right)+G(z-z';\tilde{\lambda}).
\eb
In order to prove that the formulae (\ref{eq34}) and (\ref{eq35})
indeed represent the Green function, we shall construct their
continuations to arbitrary values
$z,z'\in\Cy\backslash b$, and show that these continuations coincide with each other.

Let us start, say, from the representation (\ref{eq34}). At the first
stage, we construct its continuation to arbitrary values of $z$
only. This can be done by shifting the contours $C_-$ and $C_+$ in the
integral over $\theta$ to $\mathrm{Im}\,\theta=-\frac{\pi}{2}$,
$\mathrm{Im}\,\theta=\frac{\pi}{2}$ respectively. After this procedure
one obtains
\ben
  G^{0,\lambda}(z,z')-G(z-z';\lambda_0)=i m
  \sin\pi\lambda_1\int\limits_{-\infty}^{\infty}
 \frac{d\theta}{2\pi}\int\limits_{C_-\cup   C_+}\!\!\!\!
 \frac{d\theta'}{2\pi}\;
 \frac{G(\theta';1-\lambda)e^{\theta'+mx'\cosh\theta'+imy'\sinh\theta'}}{1-e^{im\beta\sinh\theta'+2\pi i
      \lambda_0}}\times
 \ebn
 \ben
 \times\left\{\frac{G(\theta-i\pi/2;\lambda)e^{-imx\sinh\theta+my\cosh\theta}}
 {(1-e^{m\beta\cosh\theta-2\pi i \lambda_0})(-ie^{\theta}+e^{\theta'})}\left(\begin{array}{cc}
 -ie^{\theta+\theta'} & -ie^{\theta} \\
 e^{\theta'} & 1\end{array}\right)\right.-
\ebn\ben
 \left.-\frac{G(\theta+i\pi/2;\lambda)e^{imx\sinh\theta-my\cosh\theta}}
 {(1-e^{-m\beta \cosh\theta-2\pi i \lambda_0})(ie^{\theta}+e^{\theta'})}\left(\begin{array}{cc}
 ie^{\theta+\theta'} & ie^{\theta} \\
 e^{\theta'} & 1\end{array}\right)\right\}.
 \ebn
 We can not do the same thing the second time, since the function standing
 in the integral over $\theta'$ has the poles
 $\theta'=\theta\pm\frac{i\pi}{2}$. However, one can shift $C_-$ and
 $C_+$ to the contours $C_-^{\varepsilon,\theta}$ and  $C_+^{\varepsilon,\theta}$
 shown in the Fig.~4, and then let $\varepsilon\rightarrow0$.  The
 continuation of (\ref{eq34}) to all $z,z'\in\Cy\backslash b$ is
 then
 \ben
 G^{0,\lambda}(z,z')=im\sin\pi\lambda_1\!\!\!\sum\limits_{\sigma=\pm1}\int\limits_{-\infty}^{\infty}
 \frac{d\theta}{2\pi}\int\limits_{-\infty}^{\infty}
 \frac{d\theta'}{2\pi}\;\frac{G(\theta+\frac{i\sigma\pi}{2};\lambda)G(\theta'+\frac{i\sigma\pi}{2};1-\lambda)e^{\theta'}}
 {(1-e^{-\sigma m\beta \cosh\theta-2\pi i \lambda_0})(1-e^{-\sigma m\beta \cosh\theta'+2\pi i \lambda_0})}\times
 \ebn
 \be\label{gre}
 \times \frac{e^{\sigma
     m(ix\sinh\theta-y\cosh\theta+ix'\sinh\theta'-y'\cosh\theta')}}{e^{\theta}+e^{\theta'}}
 \left(\begin{array}{cc}
 -e^{\theta+\theta'} & \sigma i e^{\theta} \\
 \sigma i e^{\theta'} & 1\end{array}\right)+
 \eb
 \ben
 -im\sin\pi\lambda_1\!\!\!\sum\limits_{\sigma=\pm1}\int\limits_{-\infty}^{\infty}
 \frac{d\theta}{2\pi}\;P\!\!\!\int\limits_{-\infty}^{\infty}
 \frac{d\theta'}{2\pi}\;\frac{G(\theta+\frac{i\sigma\pi}{2};\lambda)G(\theta'-\frac{i\sigma\pi}{2};1-\lambda)e^{\theta'}}
 {(1-e^{-\sigma m\beta \cosh\theta-2\pi i \lambda_0})(1-e^{\sigma m\beta \cosh\theta'+2\pi i \lambda_0})}\times
\ebn
 \ben
 \times \frac{e^{\sigma
     m(ix\sinh\theta-y\cosh\theta-ix'\sinh\theta'+y'\cosh\theta')}}{e^{\theta'}-e^{\theta}}
 \left(\begin{array}{cc}
 e^{\theta+\theta'} & \sigma i e^{\theta} \\
 -\sigma i e^{\theta'} & 1\end{array}\right)+\frac12\left\{G(z-z';\lambda_0)+G(z-z';\tilde{\lambda})\right\}.
 \ebn
 The last two terms are ``pole contributions'' that can be calculated
 using (\ref{rel0}) and contour representations of the Green function
 on the cylinder without branchpoints.

   \begin{figure}[h]
\begin{center}
 \includegraphics[height=4.5cm]{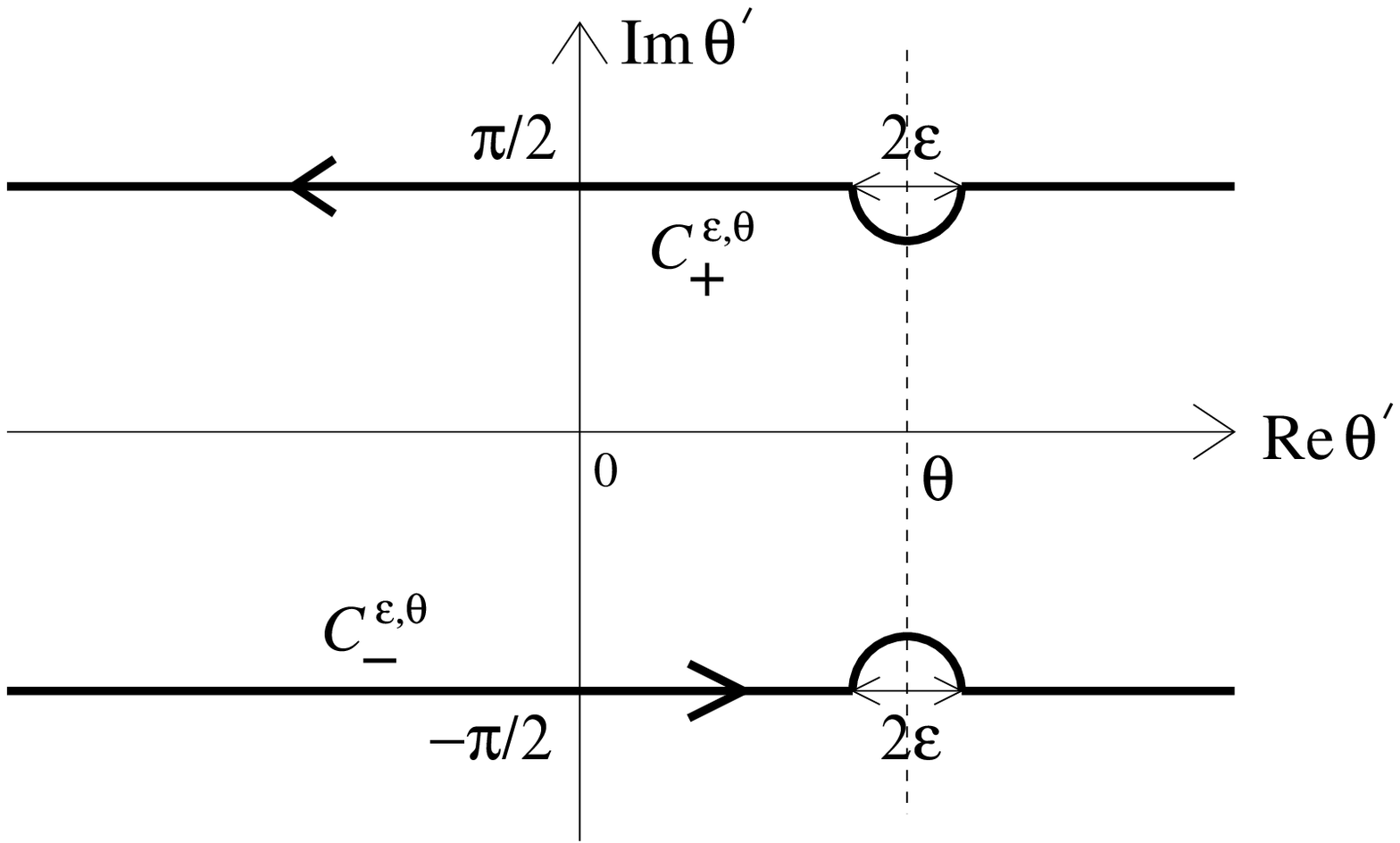}\vspace{0.5cm}

 Fig. 4. $\qquad$
 \end{center}
 \end{figure}

 If we perform the same manipulations with the representation (\ref{eq35})
 in the right half-strip, the final expression will coincide with
 (\ref{gre}) due to the relations (\ref{fr1}) and (\ref{fr2})
 satisfied by the  functions $G(\theta)$ and $H(\theta)$. Thus the formulae
 (\ref{eq34})--(\ref{gre}) indeed define the Green function
 $G^{0,\lambda}(z,z')$.

 \section{Tau functions}
 In this section we study the spaces of boundary values of some local solutions
 to the Dirac equation. These spaces can be embedded into an
 infinite-dimensional grassmannian. The $\tau$-functions are
 determined via the trivialization of the $\mathrm{det}^*$-bundle over
 this grassmannian.
 \subsection{Subspaces $W_{int}(a)$ and $W_{ext}(a)$}
 Let us consider a circle $L_{x_0}=\{(x,y)\!\!\in\!\!\Cy:x=x_0\}$, and denote
 by $H^{1/2}_{\,\lambda}(L_{x_0})$ the space of $\Cb^2$-valued quasiperiodic
 functions on $L_{x_0}$. Namely, if $g\in
 H^{1/2}_{\,\lambda}(L_{x_0})$, then
 $
 g(y+\beta)=e^{2\pi i \lambda}g(y)
 $.
 After Fourier transform the function $g$ can be written as
 \ben
 g(y)=\frac{2\pi}{\beta}\sum\limits_{n\in\Zb+\lambda}
 \hat{g}(\theta_n)e^{imy\sinh\theta_n},\qquad \sinh\theta_n=\frac{2\pi}{m\beta}\,n.
 \ebn
 Let us introduce two operators, $Q_+$ and $Q_-$, acting on
 $H^{1/2}_{\,\lambda}(L_{x_0})$ in the following way
 \ben
 Q_{\pm}g(y)=\frac{2\pi}{\beta}\sum\limits_{n\in\Zb+\lambda}
 Q_{\pm}(\theta_n)\hat{g}(\theta_n)e^{imy\sinh\theta_n},
 \ebn
 \ben
 Q_+(\theta)=\frac{1}{2\cosh\theta}\left(\begin{array}{cc}
 e^{\theta} & 1 \\
 1 & e^{-\theta}
\end{array}\right),\qquad Q_-(\theta)=\frac{1}{2\cosh\theta}\left(\begin{array}{cc}
 e^{-\theta} & -1 \\
 -1\; & \;e^{\theta}
\end{array}\right).
 \ebn
 These operators have the properties of projectors,
 \ben
 Q_++Q_-=\mathbf{1},\qquad Q_+^2=Q_+,\qquad Q_-^2=Q_-,
 \ebn
 and thus define the splitting
 $H^{1/2}_{\,\lambda}(L_{x_0})=H^+_{\lambda}\oplus H^-_{\lambda}$, with
 $H^{\pm}_{\lambda}=Q_{\pm} H^{1/2}_{\,\lambda}(L_{x_0})$. One may
 easily verify that
 \ben
 \sum\limits_{n\in\Zb+\lambda} \| Q_{\pm}(\theta_n)\hat{g}(\theta_n)\|^2
 \cosh\theta_n=\frac12 \sum\limits_{n\in\Zb+\lambda} |g_{\pm}(\theta_n)|^2,
 \ebn
 where
 \ben
 \left(\begin{array}{c}
 g_+(\theta_n) \\
 g_-(\theta_n)\end{array}\right)=\left(\begin{array}{cc}
 v_n^{1/2} & v_n^{-1/2} \\
 - v_n^{-1/2} & v_n^{1/2}\end{array}\right)\left(\begin{array}{c}
 \hat{g}_1(\theta_n) \\
 \hat{g}_2(\theta_n)\end{array}\right),\qquad v_n\equiv v(\theta_n)=e^{\theta_n}.
 \ebn
 Therefore, the function $g$ is expressed through its polarization
 components $g_{\pm}(\theta_n)$ as
 \be\label{fou}
 g(y)=\frac{2\pi}{\beta}\sum\limits_{n\in\Zb+\lambda}
 \frac{e^{imy\sinh\theta_n}}{2\cosh\theta_n}\left(\begin{array}{cc}
 v_n^{1/2} & -v_n^{-1/2} \\
 v_n^{-1/2} & v_n^{1/2}\end{array}\right)\left(\begin{array}{c}
 {g}_+(\theta_n) \\
 {g}_-(\theta_n)\end{array}\right).
 \eb

 Let us now show that the elements of $H^-_{\lambda}$
 ($H^+_{\lambda}$) represent the boundary values of quasiperiodic solutions
 to the Dirac equation in the right (left) half-strip $x>x_0$
 ($x<x_0$). To do this, rewrite the Dirac equation in the form
 \be\label{dirr}
 \partial_x\psi=\left(\begin{array}{cc}
 -i\partial_y & m \\
 m & i\partial_y \end{array}\right)\psi.
 \eb
 If we put the initial condition $\psi(x_0,y)=g(x_0,y)$ with $g(x_0,y)\in
 H^-_{\lambda}$ (i.e. all $g_+(\theta_n)=0$), the solution of
 (\ref{dirr}) in the right
 half-strip is
 \ben
 \psi_{x>x_0}=\frac{2\pi}{\beta}\sum\limits_{n\in\Zb+\lambda}
 \frac{e^{imy\sinh\theta_n-m(x-x_0)\cosh\theta_n}}{2\cosh\theta_n}\left(\begin{array}{cc}
 v_n^{1/2} & -v_n^{-1/2} \\
 v_n^{-1/2} & v_n^{1/2}\end{array}\right)\left(\begin{array}{c}
 0 \\ {g}_-(\theta_n)
 \end{array}\right).
 \ebn
 The convergence of this series is guaranteed by its convergence for
 $x=x_0$. The solution of (\ref{dirr}) in the left half-strip can be
 constructed from the element of $H^+_{\lambda}$ in a similar fashion.

 The Green function $G(z-z';\lambda)$ on the cylinder without
 branchpoints provides a useful formula for the projections $Q_{\pm}$.
 \begin{prop}
 Consider the map $Q:H^{1/2}_{\;\lambda}(L_{x_0})\rightarrow
 H^1_{\lambda}(\Cy\backslash L_{x_0})$, defined by
 \be\label{qpm}
 (Qg)(z)=i\int\nolimits_{L_{x_0}}\!\!\!G(z-z';\lambda)\,\sigma_z\,
 g(y')dy',\qquad \sigma_z=\left(\begin{array}{cc}
 1 & 0 \\
 0 & -1 \end{array}\right),\qquad g\in H^{1/2}_{\;\lambda}(L_{x_0}).
 \eb
 Then the boundary values on $L_{x_0}$ of the restrictions of $(Qg)(z)$
 to the left and right half-strip are equal to $Q_+ g$ and $-Q_-g$ respectively.
 \end{prop}
 $\blacksquare$ To prove the proposition, one has only to substitute
 in (\ref{qpm}) the Fourier expansion (\ref{fou}) of $g$ and the representations
 (\ref{green000}) and (\ref{green001}) of the Green
 function. $\blacksquare$

 Suppose no two branchpoints have the same first coordinate. Then one
 can isolate the branchcuts $b_1,\ldots,b_n$ in the open strips
 $S_1,\ldots,S_n$ (Fig. 5). The union $\bigcup\limits_{j=1}^n S_j$
 will be denoted by $S$. Consider the
 subspace of $H^1(\Cy\backslash \bar{S})$ consisting of functions $\psi$ that
 satisfy on $\Cy\backslash \bar{S}$ the Dirac equation and
 appropriate quasiperiodicity conditions:
 \ben
 \psi(x,y+\beta)=\begin{cases}
 e^{2\pi i \lambda_0}\psi(x,y)\text{ for }x<x_1^L, \\
 \exp\{2\pi i \sum\limits_{j=0}^k\lambda_j\}\psi(x,y)\text{ for
 }x_{k}^{R}<x<x_{k+1}^{L}, \\
 \exp\{2\pi i \sum\limits_{j=0}^n\lambda_j\}\psi(x,y)\text{ for }x>x_n^R.
\end{cases}
 \ebn
 We will denote by $W_{ext}$ the space of boundary values of  such
 functions. It is a subspace of $W$, the
 space of all quasiperiodic $\Cb^2$-valued $H^{1/2}$ functions on
 $\partial S$:
 \be\label{spacee}
 W=H^{1/2}_{\lambda_0}(L_{x^L_1})\oplus
 H^{1/2}_{\lambda_0+\lambda_1}(L_{x^R_1})\oplus\ldots
 \oplus H^{1/2}_{\sum\nolimits_{k=0}^n\lambda_k}(L_{x^R_n}).
 \eb
 Analogously, $W_{int}\subset W$ is defined as the space of boundary
 values of functions $g\in\mathcal{D}^{a,\lambda}$ that solve
 $D^{a,\lambda}g=0$  on $S$.
 \begin{figure}[h]
\begin{center}
 \includegraphics[height=4.5cm]{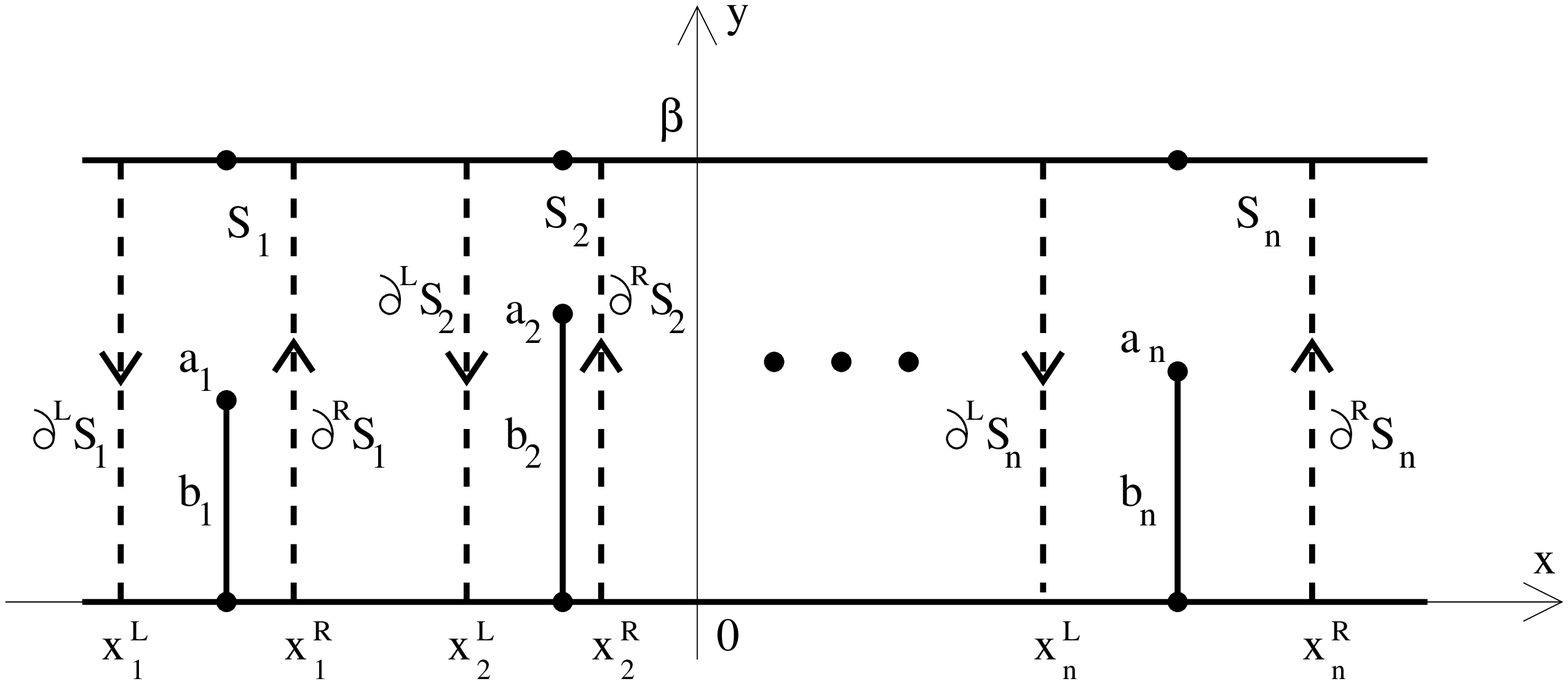}\vspace{0.5cm}

 Fig. 5. $\qquad$
 \end{center}
 \end{figure}

 The construction of the infinite-dimensional grassmannian in the next
 subsection heavily relies on the transversality of the subspaces
 $W_{ext}$ and $W_{int}$ in $W$. We postpone the proof of this fact;
 instead, let us explain how one can find the explicit
 formulae for the  projections on these subspaces. Consider the
 restriction
 $g^{(i)}=g\bigl|_{\partial^LS_i\cup\partial^RS_i}\bigr.$ of an
 element $g\in W$ to the boundary of the strip $S_i$. It is convenient
 to introduce the notation
 \ben
 g^{(i)}=\left(\begin{array}{c}g^{(i)}_{R,+}\\
     g^{(i)}_{L,-}\end{array}\right)\oplus
 \left(\begin{array}{c}g^{(i)}_{R,-}\\ g^{(i)}_{L,+}\end{array}\right).
 \ebn
 \\ \textbf{Example}. Assume for a moment that the strip $S_i$
 contains no branchcuts at all, i.~e. both $g^{(i)}_L$ and
 $g^{(i)}_R$ obey the same quasiperiodicity conditions, say,
 \ben
 g^{(i)}_L(y+\beta)=e^{2\pi i \lambda}g^{(i)}_L(y),\qquad
 g^{(i)}_R(y+\beta)=e^{2\pi i \lambda}g^{(i)}_R(y).
 \ebn
 Then the map
 \ben
 \tilde{Q}g^{(i)}(z)=i\!\!\!\!\!\!\int\limits_{\partial^LS_i\cup\partial^RS_i}
\!\!\!\!\!\! G(z-z';\lambda)\sigma_z\, g^{(i)}(y')dy'=
 \ebn
\be\label{example}
 =\int\limits_{\partial^LS_i\cup\partial^RS_i}
\!\!\!\!\!\! G_{\cdot,1}(z-z';\lambda)g^{(i)}_1(z')dz'+ G_{\cdot,2}(z-z';\lambda)g^{(i)}_2(z')d\bar{z'}
\eb
 defines a function that satisfies the Dirac equation in $S_i$. After a
 simple calculation, involving Fourier representations of $g^{(i)}$ and
 of the Green function, one obtains the explicit formula
 \ben
 \tilde{Q}g^{(i)}(z)=\frac{2\pi}{\beta}\sum\limits_{n\in\Zb+\lambda}
\frac{e^{m(x-x_i^R)\cosh\theta_n+imy\sinh\theta_n}}{2\cosh\theta_n}\left(\begin{array}{cc}
 v_n^{1/2} & -v_n^{-1/2} \\
 v_n^{-1/2} & v_n^{1/2}\end{array}\right)\left(\begin{array}{c}
  {g}_{R,+}^{(i)}(\theta_n) \\ 0
 \end{array}\right)+
 \ebn
 \ben +\frac{2\pi}{\beta}\sum\limits_{n\in\Zb+\lambda}
 \frac{e^{-m(x-x_i^L)\cosh\theta_n+imy\sinh\theta_n}}{2\cosh\theta_n}\left(\begin{array}{cc}
 v_n^{1/2} & -v_n^{-1/2} \\
 v_n^{-1/2} & v_n^{1/2}\end{array}\right)\left(\begin{array}{c}
 0 \\ {g}_{L,-}^{(i)}(\theta_n)
 \end{array}\right).
 \ebn
 Passing to boundary values, we see that $\tilde{Q}$ induces a map on
 $W$. It is given by
 \ben
 \tilde{Q}:\; \left(\begin{array}{c}g^{(i)}_{R,+}\\
     g^{(i)}_{L,-}\end{array}\right)\oplus
 \left(\begin{array}{c}g^{(i)}_{R,-}\\
     g^{(i)}_{L,+}\end{array}\right)\mapsto
 \left(\begin{array}{c}g^{(i)}_{R,+}\\
     g^{(i)}_{L,-}\end{array}\right)\oplus
  \left(\begin{array}{cc} 0 & \hat{\omega} \\
     \hat{\omega} & 0 \end{array}\right)
  \left(\begin{array}{c}g^{(i)}_{R,+}\\
     g^{(i)}_{L,-}\end{array}\right),
 \ebn
 where $(\hat{\omega} g)(\theta_n)=
 e^{-m(x^R_i-x^L_i)\cosh\theta_n}g(\theta_n)$ in Fourier representation. Furthermore, $\tilde{Q}$ is
 a projection onto the space of solutions to the Dirac
 equation on $S_i$. If $g_L^{(i)}$ and  $g_R^{(i)}$ represent boundary
 values of a function $f$ that belongs to this space, the one-form in
 (\ref{example}) is closed, so the contour of integration can be shrunk to a
 small circle around $z$. Using the asymptotics of the Green
 function at $z'\rightarrow z$, one obtains
 $\tilde{Q}f(z)=f(z)$.\vspace{0.3cm}

 The generalization of the example we have just considered to the
 strip containing a branchcut leads to
 the main technical result of this subsection:
 \begin{thm}\label{proj}  Suppose that $G^{a,\lambda}(z,z')$ is the
   one-point Green function\footnote{Here $a$ denotes a single point,
    and not the collection $(a_1,\ldots,a_n)$. I hope this abuse of
    notation  will not confuse the reader.} found in the previous
  chapter. Suppose that $a\in S'$,
  $S'=\{(x,y)\in\Cy:x_L<x<x_R\}$. Consider the function $g$ on
  $\partial S'$, which satisfies $g\bigl|_{\partial^LS'}\bigr.\in
  H^{1/2}_{\lambda_0}(\partial^LS')$, $g\bigl|_{\partial^RS'}\bigr.\in
  H^{1/2}_{\tilde{\lambda}}(\partial^RS')$. Then the map
  \be\label{proj1}
  P_{S'}(a)g(z)= \!\!\!\!\int\limits_{\partial^LS'\cup\partial^RS'}
\!\!\!\!\!\! G^{a,\lambda}_{\cdot,1}(z,z')g_1(z')dz'+
G^{a,\lambda}_{\cdot,2}(z,z')g_2(z')d\bar{z'}
  \eb
 defines a projection onto the space of functions $f\in
 \mathcal{D}^{a,\lambda}$ that  solve $D^{a,\lambda}f=0$ on $S'$. The
 induced map of boundary values is determined by the following
 formula:
 \be\label{proj2}
 P_{S'}(a):\; \left(\begin{array}{c}g_{R,+}\\
     g_{L,-}\end{array}\right)\oplus
 \left(\begin{array}{c}g_{R,-}\\
     g_{L,+}\end{array}\right)\mapsto
 \left(\begin{array}{c}g_{R,+}\\
     g_{L,-}\end{array}\right)\oplus
  \left(\begin{array}{cc} \hat{\alpha} & \hat{\beta} \\
     \hat{\gamma} & \hat{\delta} \end{array}\right)
  \left(\begin{array}{c}g_{R,+}\\
     g_{L,-}\end{array}\right),
 \eb
 where
 \begin{eqnarray}\label{alpha}
 (\hat{\alpha}
 g)(\theta_l)=\frac{2\sin\pi\lambda_1}{\beta}\sum\limits_{n\in\Zb+\tilde{\lambda}}\,
 \frac{(v_l
   v_n)^{\lambda_1+\frac12}}{1+v_l\,v_n\;}\frac{e^{-m(x_R-a_x)(\cosh\theta_l+\cosh\theta_n)-ima_y(\sinh\theta_l-\sinh\theta_n)}}
  {\cosh\theta_n}\times \\
\nonumber  \times e^{-\frac i2
   (\nu_l+\nu_n)-\frac12(\eta_l+\eta_n)}g(\theta_n),\qquad l\in\Zb+\tilde{\lambda},
 \end{eqnarray}
 \begin{eqnarray}\label{beta}
 (\hat{\beta}
 g)(\theta_l)=\frac{2e^{-\pi i\lambda_1}\sin\pi\lambda_1}{\beta}\sum\limits_{n\in\Zb+\lambda_0}\,
 \frac{e^{-m(x_R-a_x)\cosh\theta_l+m(x_L-a_x)\cosh\theta_n-ima_y(\sinh\theta_l-\sinh\theta_n)}}
  {\cosh\theta_n}\times \\
 \nonumber \times \frac{v_l^{\lambda_1+\frac12}v_n^{-\lambda_1+\frac12}}
 {v_l-v_n}e^{-\frac i2
   (\nu_l-\nu_n)-\frac12(\eta_l-\eta_n)}g(\theta_n),\qquad l\in\Zb+\tilde{\lambda},
 \end{eqnarray}
 \begin{eqnarray}\label{gamma}
 (\hat{\gamma}
 g)(\theta_l)=-\frac{2e^{\pi i\lambda_1}\sin\pi\lambda_1}{\beta}\sum\limits_{n\in\Zb+\tilde{\lambda}}\,
 \frac{e^{m(x_L-a_x)\cosh\theta_l-m(x_R-a_x)\cosh\theta_n-ima_y(\sinh\theta_l-\sinh\theta_n)}}
  {\cosh\theta_n}\times
 \\ \nonumber \times \frac{v_l^{-\lambda_1+\frac12}v_n^{\lambda_1+\frac12}}
 {v_l-v_n}e^{\frac i2
   (\nu_l-\nu_n)+\frac12(\eta_l-\eta_n)}g(\theta_n),\qquad l\in\Zb+\lambda_0,
 \end{eqnarray}
 \begin{eqnarray}\label{delta}
 (\hat{\delta}
 g)(\theta_l)=\frac{2\sin\pi\lambda_1}{\beta}\sum\limits_{n\in\Zb+\lambda_0}\,
 \frac{(v_l
   v_n)^{-\lambda_1+\frac12}}{1+v_l\,v_n\;}\frac{e^{m(x_L-a_x)(\cosh\theta_l+\cosh\theta_n)-ima_y(\sinh\theta_l-\sinh\theta_n)}}
  {\cosh\theta_n}\times
 \\ \nonumber \times e^{\frac i2
   (\nu_l+\nu_n)+\frac12(\eta_l+\eta_n)}g(\theta_n),\qquad l\in\Zb+\lambda_0,
 \end{eqnarray}
 and $\nu_l=\nu(\theta_l;\lambda)$, $\eta_l=\eta(\theta_l;\lambda)$.
 \end{thm}
 $\blacksquare$ The derivation of (\ref{proj})--(\ref{delta}) can be
 carried out analogously to the previous example, using two more (in
 addition to (\ref{eq33}--\ref{eq35}))
 representations of the one-point Green function:
  \be\label{zazp}
 G^{a,\lambda}(z,z')=i\sin\pi\lambda_1\sum\limits_{l\in\Zb+\lambda_0}\sum\limits_{n\in\Zb+\tilde{\lambda}}
 \frac{G(\theta_l;\lambda)H(\theta_n;1-\lambda)}
  {m\beta^2\cosh\theta_l\cosh\theta_n}\,\frac{e^{\theta_n+m(x-a_x)\cosh\theta_l+i
   m(y-a_y)\sinh\theta_l}}{e^{\theta_l}-e^{\theta_n}}\times
\eb
 \ben
 \times e^{ - m(x'-a_x)\cosh\theta_n
 -im(y'-a_y)\sinh\theta_n} \left(\begin{array}{cc}
 -e^{\theta_l+\theta_n} & e^{\theta_l} \\
  -e^{\theta_n} & 1\end{array}\right)\quad\text{ for } x<a_x<x',
\ebn
 \be\label{zpaz}
 G^{a,\lambda}(z,z')=i\sin\pi\lambda_1\sum\limits_{l\in\Zb+\lambda_0}\sum\limits_{n\in\Zb+\tilde{\lambda}}
 \frac{G(-\theta_l;\lambda)H(-\theta_n;1-\lambda)}
  {m\beta^2\cosh\theta_l\cosh\theta_n}\,\frac{e^{-\theta_n-m(x-a_x)\cosh\theta_l+i
   m(y-a_y)\sinh\theta_l}}{e^{-\theta_l}-e^{-\theta_n}}\times
\eb
 \ben
 \times e^{ m(x'-a_x)\cosh\theta_n
 -im(y'-a_y)\sinh\theta_n} \left(\begin{array}{cc}
 -e^{-\theta_l-\theta_n} & -e^{-\theta_l} \\
  e^{-\theta_n} & 1\end{array}\right)\quad\text{ for } x>a_x>x'.
\ebn
 When one applies the Stokes theorem to prove the projection property,
 the contour of integration in (\ref{proj1}) can be deformed into two
 small circles, around $z$ and $a$.  Using the expansions
 (\ref{h1}), (\ref{green2}) of the
 multivalued local solution to Dirac equation and Green function, one  easily shows
 that the second integral vanishes.
 $\blacksquare$
 \\ \textbf{Remark}. Let us choose in $H^{1/2}_{\lambda}(L)$ a
 complete orthonormal family $\{\varphi_k\}$, say,
 \ben
 \varphi_k=\frac{1}{\sqrt{\beta}}\, e^{i m y\sinh\theta_k},\quad k\in \Zb+\lambda.
 \ebn
 With these functions, we can find Schmidt norms of $\hat{\alpha}$, $\hat{\beta}$,
$\hat{\gamma}$, $\hat{\delta}$ and show that they are finite. For
example,
 \ben
 \|\hat{\beta}\|_2^{\,2}=\sum\limits_{n,n'\in\Zb}|\langle\hat{\beta}\varphi_{n+\lambda_0},\varphi_{n'+\tilde{\lambda}}\rangle|^2
 =\sum\limits_{l\in\Zb+\tilde{\lambda}}\sum\limits_{n\in\Zb+\lambda_0}|\hat{\beta}(\theta_l,\theta_n)|^2,
 \ebn
 where $\hat{\beta}(\theta_l,\theta_n)$  denotes the ``kernel'' of
 $\hat{\beta}$. This sum rapidly converges due to the exponential factors
 $e^{-m(x_R-a_x)\cosh\theta_l}$ and $e^{m(x_L-a_x)\cosh\theta_n}$ in
 $\hat{\beta}(\theta_l,\theta_n)$. Note, however, that in the limit
 $\beta\rightarrow\infty$, when $\hat{\alpha}$, $\hat{\beta}$,
$\hat{\gamma}$, $\hat{\delta}$ become integral operators,
$\hat{\beta}$ and
$\hat{\gamma}$ no longer belong to the Schmidt class due to the
singularities in the kernels.\vspace{0.3cm}

 We briefly outline the proof of the transversality of the subspaces
 $W_{int}$ and $W_{ext}$ in $W$, closely following Palmer's work
 \cite{pacific}. Suppose that $f\in W$ is decomposed as $f=g+h$, with
 $g\in W_{int}$ and $h\in W_{ext}$. Theorem~\ref{teo1} guarantees the
 uniqueness of this decomposition, since the elements of $W_{int}\cap
 W_{ext}$ represent the boundary values of the functions from
 $\widetilde{\mathbf{W}}^{a,\lambda}$. It remains to prove only the existence.

 In order for $g$ to be in $W_{int}$, one should satisfy the
 conditions
 \be\label{relg}
  \left(\begin{array}{c}g^{(i)}_{R,-}\\
     g^{(i)}_{L,+}\end{array}\right)=\left(\begin{array}{cc} \hat{\alpha}_i & \hat{\beta}_i \\
     \hat{\gamma}_i & \hat{\delta}_i \end{array}\right)
  \left(\begin{array}{c}g^{(i)}_{R,+}\\
     g^{(i)}_{L,-}\end{array}\right),
 \eb
 where  $\hat{\alpha}_i$, $\hat{\beta}_i$,
$\hat{\gamma}_i$, $\hat{\delta}_i$ are obtained from $\hat{\alpha}$, $\hat{\beta}$,
$\hat{\gamma}$, $\hat{\delta}$ by the substitution
 \ben
x_L\rightarrow x^L_i,\quad x_R\rightarrow x^R_i,\quad a\rightarrow
a_i, \quad \lambda_0\rightarrow\sum\limits_{k=0}^{i-1}\lambda_k,\quad
\tilde{\lambda}\rightarrow \sum\limits_{k=0}^{i}\lambda_k.
 \ebn
 Another set of relations follows from the assumption that $h\in
 W_{ext}$. Indeed, $h^{(1)}_L$ should represent the boundary value of
 a solution to  Dirac equation in the left half-strip $x<x_1^L$;
 analogously, $h^{(n)}_R$ is the boundary value of a solution in the
 half-strip $x>x^R_n$. This leads to two relations,
 \be\label{relh1}
 h^{(1)}_{L,-}=0,\qquad h^{(n)}_{R,+}=0.
 \eb
 Next, the example we have considered shows that the functions $h^{(i)}_R$ and
 $h^{(i+1)}_L$ are boundary values of a solution in
 $x^R_i<x<x^L_{i+1}$, if
 \be\label{relh2}
 h^{(i+1)}_{L,-}=\hat{\omega}_i h^{(i)}_{R,-},\qquad
 h^{(i)}_{R,+}=\hat{\omega}_i h^{(i+1)}_{L,+},
 \qquad i=1,\ldots, n-1,
 \eb
 where $\hat{\omega}_i$ is obtained from $\hat{\omega}$ by the
 substitution
   \ben
   x^L_i\rightarrow x^R_i,\qquad x^R_i\rightarrow
   x^L_{i+1},\qquad \lambda\rightarrow \sum\limits_{k=0}^i\lambda_k.
   \ebn
 One can transform (\ref{relh1}) and (\ref{relh2}) into the conditions
 on $g$. Using (\ref{relg}) to eliminate all $g^{(i)}_{R,-}$ and
 $g^{(i)}_{L,+}$, we obtain a system of equations,
  \be\label{syst1}
  g^{(i)}_{R,+}-\hat{\omega}_i\left(\hat{\gamma}_{i+1}
  g^{(i+1)}_{R,+}+\hat{\delta}_{i+1}
  g^{(i+1)}_{L,-}\right)=f^{(i)}_{R,+}-\hat{\omega}_i f^{(i+1)}_{L,+},\qquad i=1,\ldots,n-1,
  \eb
  \be\label{syst2}
  g^{(i+1)}_{L,-}-\hat{\omega}_i\left(\hat{\alpha}_i
  g^{(i)}_{R,+}+\hat{\beta}_i
  g^{(i)}_{L,-}\right)=f^{(i+1)}_{L,-}-\hat{\omega}_i
f^{(i)}_{R,-},\qquad i=1,\ldots,n-1,\eb
  \be\label{syst3}
 g^{(1)}_{L,-}= f^{(1)}_{L,-},\qquad g^{(n)}_{R,+}= f^{(n)}_{R,+}.
 \eb
 If we introduce the notation
 \ben
 U_i=-\left(\begin{array}{cc}\hat{\omega}_i\hat{\gamma}_{i+1} &
     \hat{\omega}_i\hat{\delta}_{i+1} \\
 0 & 0\end{array}\right),\qquad V_i=-\left(\begin{array}{cc}0 & 0 \\ \hat{\omega}_i\hat{\alpha}_{i} &
     \hat{\omega}_i\hat{\beta}_{i} \end{array}\right), \qquad i=1,\ldots,n-1,
  \ebn
 \ben
 \tilde{g}_j=\left(\begin{array}{c}g^{(j)}_{R,+}\\
     g^{(j)}_{L,-}\end{array}\right),\qquad
 F_k=\left(\begin{array}{c}f^{(k)}_{R,+}-\hat{\omega}_k
     f^{(k+1)}_{L,+} \\
     f^{(k)}_{L,-}-\hat{\omega}_{k-1}
     f^{(k-1)}_{R,-}\end{array}\right),\qquad j=1,\ldots,n,\quad k=2,\ldots,n-1,
\ebn
\ben
 F_1=\left(\begin{array}{c}f^{(1)}_{R,+}-\hat{\omega}_1
     f^{(2)}_{L,+} \\
     f^{(1)}_{L,-}\end{array}\right),\qquad
 F_n=\left(\begin{array}{c}f^{(n)}_{R,+} \\
     f^{(n)}_{L,-}-\hat{\omega}_{n-1}
     f^{(n-1)}_{R,-}\end{array}\right),
\ebn
 the system (\ref{syst1})--(\ref{syst3}) can be rewritten as
\be\label{transv}
 \left(\begin{array}{ccccc}
 \mathbf{1} & U_1 & 0 & .  & 0 \\
 V_1 & \mathbf{1} & U_2 & . & 0 \\
 0 & V_2 & \mathbf{1} & . & . \\
 . & . & . & . & U_{n-1} \\
0 & 0 & . & V_{n-1} & \mathbf{1} \end{array}\right)
 \left(\begin{array}{c}
 \tilde{g}_1 \\ . \\ . \\ . \\ \tilde{g}_n\end{array}\right)=
 \left(\begin{array}{c}
 F_1 \\ . \\ . \\ . \\ F_n\end{array}\right).
\eb
 The operator, standing in the LHS of (\ref{transv}), represents a
 compact perturbation of the identity, and thus is Fredholm of index
 zero. Since to every nontrivial element of its kernel there corresponds a
 nontrivial element of $W_{int}\cap W_{ext}$, the kernel should be
 zero. Therefore, this operator is invertible and can be used to
 construct  the decomposition $f=g+h$ explicitly.
 \subsection{Grassmannian, $\mathrm{det}^*$-bundle and its
   trivialization}
 First we introduce several important definitions, following Segal and
 Wilson \cite{segal} (further details can be found in
 \cite{pressley,witten}).
 \begin{defin} Suppose we have a complex Hilbert space $H$ with a
   given decomposition $H=H_+\oplus H_-$.  The Grassmannian $Gr(H)$ is
   a set of all closed subspaces $V\subset H$ such that
 \begin{itemize}
 \item the projection  $\mathrm{pr_+}:V\rightarrow H_+$ along $H_-$ is a
   Fredholm operator;
 \item the projection  $\mathrm{pr_-}:V\rightarrow H_-$ along $H_+$ is a
   Hilbert-Schmidt operator.
 \end{itemize}
 \end{defin}
 The first requirement means that the codimension of $V\cap H_+$ is
 finite in both $V$ and $H_+$. The connected components of the
 Grassmannian are distinguished by the value of the index of
 $\mathrm{pr_+}$. We shall work only with the component $Gr_0(H)$ that corresponds
 to zero index.
\begin{defin}
The invertible linear map $v: H_+\rightarrow V$ is called an
admissible frame for the subspace $V\in Gr_0(H)$ if $\;\mathrm{pr_+}\circ v:
H_+\rightarrow H_+$ is a trace class perturbation of the
identity.
 The fiber of the $\mathrm{det}^*$-bundle over $V$ consists of the
equivalence classes of pairs $(v,\alpha)$, where $v$ is an admissible
frame, $\alpha$ is a complex
number and $(v_1,\alpha_1)\sim(v_2,\alpha_2)$ if
$\;\alpha_1=\alpha_2\,\mathrm{det}\left(v_2^{-1}v_1\right)$.
The canonical section  of the $\mathrm{det}^*$-bundle is given
by $\sigma:V\mapsto \bigl(v,\mathrm{det}\,(\mathrm{pr_+}\circ v)\bigr)$.
\end{defin}

In the work of Segal and Wilson, as well as in almost all subsequent
papers on the subject, the Hilbert space $H$ is the space of all
square-integrable complex-valued functions on the unit circle
$S^1=\{z\in\Cb: |z|=1\}$. $H_+$ and $H_-$ are spanned by the elements
$\{z^k\}$ with $k\geq 0$ and $k<0$, respectively.

We are interested in a more complicated  model of the
Grassmannian. $H$ is identified with the space $W$ (see
(\ref{spacee})) of square-integrable quasiperiodic $\Cb^2$-valued functions on the
boundary $\partial S$.  Let us fix a collection of points
$a^0=(a^0_1,\ldots,a^0_n)$ such that $a^0_j\in S_j$,
$j=1,\ldots,n$. Then one can define the Grassmannian $Gr(W)$ with
respect to the splitting $W=W_{int}(a^0)\oplus W_{ext}$. The
crucial observation, similar to one made by Palmer in \cite{pacific},
is that $W_{int}(a)\in Gr_0(W)$.

 Now introduce two admissible frames for the subspace $W_{int}(a)$.
The first one, which will be denoted as $P(a): W_{int}(a^0)\rightarrow
W_{int}(a)$, represents the projection of $W_{int}(a^0)$ on  $W_{int}(a)$ along
$W_{ext}$. It is easy to guess that $P(a)$ inverts
$\mathrm{pr_+}$. Thus the canonical section can be written as
$\sigma:W_{int}(a)\mapsto (P(a),1)$. The second admissible frame, $F(a): W_{int}(a^0)\rightarrow
W_{int}(a)$, is the restriction to $W_{int}(a^0)$ of the direct sum of
the appropriate one-point projections (\ref{proj1})-(\ref{proj2}):
$F(a)=\bigl(P_{S_1}(a_1)\oplus\ldots\oplus
P_{S_n}(a_n)\bigr)\Bigl|_{W_{int}(a^0)}\Bigr.$. It defines a second,
trivializing section $\vartheta:W_{int}(a)\mapsto (F(a),1)$. The
determinant of the Dirac operator $D^{a,\lambda}$, or the
$\tau$-function, is determined from the comparison of the two
sections,
\be\label{tauf}
\tau(a,a^0)=\frac{\sigma(W_{int}(a))}{\vartheta(W_{int}(a))}=\mathrm{det}\left(P(a)^{-1}F(a)\right).
\eb
\textbf{Remark}. This ideology originates from the work
\cite{cauchy}, where the isomonodromic $\tau$-function, associated to
a fuchsian system on $\Cb\Pb^1$, was interpreted as the determinant of
a Cauchy-Riemann operator, whose domain incorporates functions with
specified branching.\vspace{0,3cm}

In order to calculate $\tau(a,a^0)$ more explicitly, let us use
several results and notations from the previous subsection. Suppose
that $f\in W_{int}(a^0)$, $g\in W_{int}(a)$, then
 \be\label{condit}
f^{(j)}=\tilde{f}_j\oplus N_j\left(a^0\right)\tilde{f}_j,\quad
g^{(j)}=\tilde{g}_j\oplus N_j(a)\,\tilde{g}_j,
\qquad N_j(a)=\left(\begin{array}{cc}
 \hat{\alpha}_j(a) & \hat{\beta}_j(a) \\
     \hat{\gamma}_j(a) & \hat{\delta}_j(a)
\end{array}\right).
 \eb
 The functions $f$ and $g$ can be represented by columns
 \ben
 f=(\tilde{f}_1\ldots\tilde{f}_n)^T,\qquad g=(\tilde{g}_1\ldots\tilde{g}_n)^T.
\ebn
 The map $F(a):W_{int}(a^0)\rightarrow
 W_{int}(a)$ is given in these coordinates by the identity
 transformation. To obtain the representation of $P(a)^{-1}$, for each
 $g\in W_{int}(a)$ one should find a function $f\in W_{int}(a^0)$ such
 that $g=f-h$ with $h\in W_{ext}$. This amounts to the same
 calculation as we have done earlier in
 (\ref{relh1})--(\ref{transv}). Taking into account the
 additional condition (\ref{condit}) on $f$, one finally obtains
 \ben
 (\mathbf{1}+M(a))\,g=\left(\mathbf{1}+M(a^0)\right)f,
 \ebn
 where
 \ben
 M(a)=\left(\begin{array}{ccccc}
 0 & U_1(a) & 0 & .  & 0 \\
 V_1(a) & 0 & U_2(a) & . & 0 \\
 0 & V_2(a) & 0 & . & . \\
 . & . & . & . & U_{n-1}(a) \\
0 & 0 & . & V_{n-1}(a) & 0 \end{array}\right) .
 \ebn
 Therefore, the $\tau$-function is equal to
 \be\label{tauf2}
 \tau(a,a^0)=\mathrm{det}\left\{(\mathbf{1}+M(a))\left(\mathbf{1}+M(a^0)\right)^{-1}\right\}.
 \eb
  In fact, one can derive an even more convenient representation.  Let us
 introduce the matrix
  \ben
 \tilde{M}(a)=\left(\begin{array}{ccccc}
 0 & \tilde{U}_1(a) & 0 & .  & 0 \\
 \tilde{V}_1(a) & 0 & \tilde{U}_2(a) & . & 0 \\
 0 & \tilde{V}_2(a) & 0 & . & . \\
 . & . & . & . & \tilde{U}_{n-1}(a) \\
0 & 0 & . & \tilde{V}_{n-1}(a) & 0 \end{array}\right)
 \ebn
 with
 \ben
 \tilde{U}_j=\left(\begin{array}{cc}
 -\hat{\omega}_j\hat{\gamma}_{j+1} & 0 \\ 0 & 0 \end{array}\right), \quad\tilde{V}_j=\left(\begin{array}{cc}
 0 & 0 \\ 0 & -\hat{\omega}_j\hat{\beta}_{j} \end{array}\right),\qquad
j=1,\ldots,n-1.
\ebn
 The matrix $\mathbf{1}+\tilde{M}(a)$ is a product of an upper
 triangular and a lower triangular matrix with identities on the
 diagonals. Thus we have
 \ben
 \mathrm{det}\left\{\left(\mathbf{1}+\tilde{M}(a^0)\right)\left(\mathbf{1}+\tilde{M}(a)\right)^{-1}\right\}=1.
 \ebn
 Multiplying the RHS of (\ref{tauf2}) by this determinant, one finds
 \ben
 \tau(a,a^0)=\mathrm{det}\left\{\bigl(\mathbf{1}+\tilde{M}(a)\bigr)^{-1}
 (\mathbf{1}+M(a))\left(\mathbf{1}+M(a^0)\right)^{-1}\bigl(\mathbf{1}+\tilde{M}(a^0)\bigr)\right\}=\frac{\tau(a)}{\tau(a^0)},
 \ebn
 where
 \be\label{tauf3}
 \tau(a)=\mathrm{det}\left\{\bigl(\mathbf{1}+\tilde{M}(a)\bigr)^{-1}
 (\mathbf{1}+M(a))\right\}.
 \eb
 \textbf{Example}. Consider the simplest nontrivial situation, when
 there are only two branching points on the cylinder. Let us introduce
 the notation
 $$\tilde{\lambda}=\lambda_0+\lambda_1,\quad
 \bar{\lambda}=\lambda_0+\lambda_1+\lambda_2,\qquad a_x=(a_2)_x-(a_1)_x,\quad
 a_y=(a_2)_y-(a_1)_y.$$
 In the case $n=2$ the inverse matrix
 $\bigl(\mathbf{1}+\tilde{M}(a)\bigr)^{-1}$ looks  particularly simple,
 \ben
 \bigl(\mathbf{1}+\tilde{M}(a)\bigr)^{-1}=\left(\begin{array}{cc}\mathbf{1} & -\tilde{U}_1(a) \\
 -\tilde{V}_1(a) & \mathbf{1} \end{array}\right).
 \ebn
 Using this formula, one can find
 that the two-point $\tau$-function is given by
 \be\label{taun2}
 \tau(a)=\mathrm{det}\left(\mathbf{1}-{K}\right),
 \eb
 where the operator
 ${K}=\hat{\omega}_1\hat{\alpha}_1\hat{\omega}_1\hat{\delta}_2$
 can be represented by the infinite-dimensional matrix with the entries
 \be\label{taun3}
 {K}_{mn}=\frac{4\sin\pi\lambda_1\sin\pi\lambda_2}{\beta^2}\,
 \frac{(v_mv_n)^{\frac{\lambda_1-\lambda_2+1}{2}}}{\sqrt{\cosh\theta_m\cosh\theta_n}}\times\eb\ben\times
 \sum\limits_{l\in\Zb+\tilde{\lambda}}\frac{
  v_l^{\lambda_1-\lambda_2+1} e^{-m|a_x|\frac{\cosh\theta_m+2\cosh\theta_l+\cosh\theta_n}{2}+
 ima_y\frac{\sinh\theta_m-2\sinh\theta_l+\sinh\theta_n}{2}+\frac{\rho(\theta_m)+2\rho(\theta_l)+\rho(\theta_n)}{2}}}
 {(1+v_m v_l)(1+v_lv_n)\cosh\theta_l}.
 \ebn
 The indices take on the values $ m,n\in\Zb+\tilde{\lambda}$ and
 \ben
2\rho(\theta)={\eta(\theta;\tilde{\lambda},\bar{\lambda})-\eta(\theta;\lambda_0,\tilde{\lambda})+
 i\nu(\theta;\tilde{\lambda},\bar{\lambda})-i\nu(\theta;\lambda_0,\tilde{\lambda})}.
\ebn
 One can also write $K$ as
 \ben
K=4\sin\pi\lambda_1\sin\pi\lambda_2\cdot VV^T,
\ebn
 \ben
 V_{mn}=\frac{1}{\beta}\,\frac{(v_mv_n)^{\frac{\lambda_1-\lambda_2+1}{2}}e^{-m|a_x|\frac{\cosh\theta_m+\cosh\theta_n}{2}+
 ima_y\frac{\sinh\theta_m-\sinh\theta_n}{2}+\frac{\rho(\theta_m)+\rho(\theta_n)}{2}}}{\sqrt{\cosh\theta_m\cosh\theta_n}(1+v_mv_n)},\qquad
 m,n\in\Zb+\tilde{\lambda}.
 \ebn
 These explicit formulae for the $\tau$-function are in some sense a
 reward for the technical work put in the calculation of the element
 of canonical basis on the 1-punctured cylinder
 (Theorem~\ref{teo2}). It would be interesting to compare them with
 the correlation functions of twisted fields,
 calculated in the lattice regularization of the Dirac theory on
 the plane \cite{busha}. \vspace{0,3cm}

 Remark that the final answer (\ref{taun2})--(\ref{taun3}) for the
 two-point $\tau$-function is independent of the choice of
 localization (coordinates of edges of the strips
 $S_1,\ldots,S_n$). To show that this is true in the general case, we prove
 \begin{prop}
 The logarithmic derivatives of the $\tau$-function (\ref{tauf}) are given
 by
 \be\label{dertau}
 d\ln\tau(a,a^0)=\frac{m}{2}\sum\limits_{\nu=1}^n\left\{a^{(\nu)}_{1/2}(\tilde{\mathbf{w}}_{\nu}(\lambda))\,da_{\nu}+
 \overline{a^{(\nu)}_{1/2}(\tilde{\mathbf{w}}_{\nu}(-\lambda))}\;d\bar{a}_{\nu}\right \}.
 \eb
 \end{prop}
 $\blacksquare$ Consider the $n$-point Green function
 $G^{a,\lambda}(z,z')$ and construct the map
 $\tilde{P}(a):W\rightarrow W_{int}(a)$ in the following way:
 \be\label{au}
 \tilde{P}(a)f(z)=\int\limits_{\partial S}
 G^{a,\lambda}_{\cdot,1}(z,z')f_1(z')dz'+
G^{a,\lambda}_{\cdot,2}(z,z')f_2(z')d\bar{z'}.
 \eb
 It is easy to see that this map defines the projection on
 $W_{int}(a)$ along $W_{ext}$. Indeed, let us write the function
 $f(z)$ as $f=g+h$, with $g\in W_{int}(a)$ and $h\in
 W_{ext}$. The form in the integral $\tilde{P}(a)g(z)$ is closed, thus
   each piece $\partial S_{\mu}$ of the  integration contour can be
   shrunk up to two small circles, around $z'=z$ and $z'=a_{\mu}$. Computing
   the residues, we obtain $\tilde{P}(a)g(z)=g(z)$. In a similar
     fashion one also shows that $\tilde{P}(a)h(z)=0$.

 It is clear that the admissible frame $P(a):W_{int}(a^0)\rightarrow
 W_{int}(a)$ and the projection $\mathrm{pr_+}:W_{int}(a)\rightarrow
 W_{int}(a^0)$ are the restrictions
  \ben
 P(a)=\tilde{P}(a)\Bigl|_{W_{int}(a^0)}\Bigr.,\qquad  \mathrm{pr_+}=\tilde{P}(a^0)\Bigl|_{W_{int}(a)}\Bigr..
  \ebn

 Let us analogously consider the map $\tilde{F}(a):W\rightarrow
 W_{int}(a)$, which is by definition the direct sum of the one-point
 projections,
 $ \tilde{F}(a)=P_{S_1}(a_1)\oplus\ldots\oplus P_{S_n}(a_n) $. The
 second admissible frame that we have used, $F(a):
 W_{int}(a^0)\rightarrow W_{int}(a)$, is the restriction
 $F(a)=\tilde{F}(a)\Bigl|_{W_{int}(a^0)}\Bigr.$. Its inverse, which we
 denote as $F(a^0):W_{int}(a)\rightarrow W_{int}(a^0)$, is given by
 $F(a^0)=\tilde{F}(a^0)\Bigl|_{W_{int}(a)}\Bigr.$ .

 Therefore, differentiating (\ref{tauf}), one obtains
 \ben
 d\ln\tau(a,a^0)=-\mathrm{Tr}\left\{d\left(F(a)^{-1}P(a)\right)P(a)^{-1}F(a)\right\}=-\mathrm{Tr}\left\{F(a^0)
 d(P(a))\mathrm{pr_+}F(a)\right\}.
 \ebn
 Recall that the traces in the last formula, and the
 determinant in (\ref{tauf}) as well, are calculated on the subspace
 $W_{int}(a^0)$. However, since the range of both $F(a^0)$ and $\tilde{F}(a^0)$ is
 $W_{int}(a^0)$, we can forget about this restriction and replace under the
 last trace $P(a)$ by
 $\tilde{P}(a)$, $F(a)$ by $\tilde{F}(a)$, $\mathrm{pr_+}$ by
 $\tilde{P}(a^0) $ and $F(a^0)$ by $\tilde{F}(a^0)$. If we also use
 the relations
 \ben \tilde{P}(a)(\mathbf{1}-\tilde{P}(a^0))=0,\qquad
 \tilde{F}(a)(\mathbf{1}-\tilde{F}(a^0))=0,\ebn the trace becomes
  \ben
 d\ln\tau(a,a^0)=-\mathrm{Tr}_{_{W}}\!\!\left\{\tilde{F}(a^0)
 d(\tilde{P}(a))\tilde{P}(a^0)\tilde{F}(a)\right\}=-\mathrm{Tr}_{_{W}}\!\!\left\{d(\tilde{P}(a))\tilde{F}(a)\right\}.
 \ebn

 Taking into account the explicit form (\ref{au}) of $\tilde{P}(a)$
 and the formulae (\ref{der1})--(\ref{der2}) for the derivatives of
 the Green function, one can show that $d\tilde{P}(a)$ is an integral
 operator with degenerate kernel. Then we have, for example,
 \ben
 {\partial_{
     a_{\nu}}}\ln\tau(a,a^0)=\frac{im^2}{8\sin\pi\lambda_{\nu}}\sum\limits_{\mu=1}^n\;\int\limits_{\partial S_{\mu}}
 \left\{\Bigl(\tilde{\mathbf{w}}_{\nu}(z,-\lambda)\Bigr)_1\Bigl(P_{S_{\mu}}(a_{\mu})\tilde{\mathbf{w}}_{\nu}(z,\lambda)\Bigr)_1 dz+\right.\ebn
 \ben +
 \left. \Bigl(\tilde{\mathbf{w}}_{\nu}(z,-\lambda)\Bigr)_2\Bigl(P_{S_{\mu}}(a_{\mu})\tilde{\mathbf{w}}_{\nu}(z,\lambda)\Bigr)_2 d\bz\right\}.
 \ebn
 Again applying the Stokes theorem, each contour $\partial S_{\mu}$
 can be shrunk to a small circle around $a_{\mu}$. Only one circle,
 around $a_{\nu}$, gives a non-zero contribution, which can be
 calculated using the asymptotics of the one-point Green function. At
 the end of this calculation one finds
 \ben {\partial_{
     a_{\nu}}}\ln\tau(a,a^0)=\frac m2\,a^{(\nu)}_{1/2}(\tilde{\mathbf{w}}_{\nu}(\lambda)),\qquad {\partial_{
     \bar{a}_{\nu}}}\ln\tau(a,a^0)=\frac m2\,\overline{a^{(\nu)}_{1/2}(\tilde{\mathbf{w}}_{\nu}(-\lambda))},\ebn
 as claimed.  $\blacksquare$

 \section{Deformation equations}
 Let us now find the differential equations satisfied by the elements
 (\ref{basis2}), using the idea that we have already exploited in the
 calculation of the derivative of the Green function. For example, consider the
 solution $\tilde{\mathbf{w}}_{\mu}(\lambda)$ and differentiate it with
 respect to $a_{\rho}$. We obtain again a solution of the Dirac
 equation but with more singular local expansions at the branchpoints:
 $$\partial_{a_{\rho}}\tilde{\mathbf{w}}_{\mu}(\lambda)[a_{\nu}]=\sum\limits_{k>0}
 \left\{\partial_{a_{\rho}}a^{(\nu)}_{\,k}(\tilde{\mathbf{w}}_{\mu}(\lambda))\;w_{k+\lambda_{\nu}}[a_{\nu}]+
 \partial_{a_{\rho}}b^{(\nu)}_{\,k}(\tilde{\mathbf{w}}_{\mu}(\lambda))\;w^{\;*}_{k-\lambda_{\nu}}[a_{\nu}]\right\}-$$
$$-\frac m2 \,\delta_{\rho \nu}\left[
  \delta_{\mu
     \nu}w_{-3/2+\lambda_{\nu}}[a_{\nu}]+
 \sum\limits_{k>0}
 \left\{a^{(\nu)}_{\,k}(\tilde{\mathbf{w}}_{\mu}(\lambda))\;w_{k-1+\lambda_{\nu}}[a_{\nu}]+
 b^{(\nu)}_{\,k}(\tilde{\mathbf{w}}_{\mu}(\lambda))\;w^{\;*}_{k+1-\lambda_{\nu}}[a_{\nu}]\right\}\right].$$
 Adding appropriate linear combination of
 $\{\tilde{\mathbf{w}}_{\eta}(\lambda)\}$,
 $\{\partial_z\tilde{\mathbf{w}}_{\eta}(\lambda)\}$ and $\{\partial_{\bz}\tilde{\mathbf{w}}_{\eta}(\lambda)\}$
 ($\eta=1,\ldots,n$), one can annihilate the coefficients near the
 ``extra'' terms $w_{-3/2+\lambda_{\nu}}$ and
 $w_{-1/2+\lambda_{\nu}}$. Then the result will vanish identically,
 since it is clearly in
 $\widetilde{\mathbf{W}}^{a,\lambda}$. This observation can be written
 in the following general form:
 \be\label{holsys}
 d_{a,\bar{a}} \vec{\mathbf{w}}(\lambda)=(\Phi\,\partial_z+\Phi^*\partial_{\bz}+\Psi)\,\vec{\mathbf{w}}(\lambda).
 \eb
 Here
 $d_{a,\bar{a}}=\sum\limits_{j=1}^n(da_j\cdot\partial_{a_j}+d\bar{a}_j\cdot\partial_{\bar{a}_j})$ denotes the differential with
 respect to the positions of the singularities, $\Phi$, $\Phi^*$ and
 $\Psi$ are matrix-valued one-forms, and
 $\vec{\mathbf{w}}(\lambda)=(\tilde{\mathbf{w}}_{1}(\lambda)\ldots \tilde{\mathbf{w}}_{n}(\lambda))^T$.

 Let us introduce the notation
 \be\label{ccet}
 C_j=\left[a_{j+1/2}^{(\nu)}(\tilde{\mathbf{w}}_{\mu}(\lambda))\right]_{\mu,\nu=1,\ldots,n},\quad
 C^{\,*}_j=\left[b_{j-1/2}^{(\nu)}(\tilde{\mathbf{w}}_{\mu}(\lambda))\right]_{\mu,\nu=1,\ldots,n},\qquad
 j\in\Zb.
 \eb
 In particular, one has $C_0=\mathbf{1}$ and $C_j=C_j^{\,*}=\mathbf{0}$
 for $j<0$. The system (\ref{holsys}), being rewritten in terms of
 $\{C_j\}$, $\{C_j^{\,*}\}$, amounts to
 \be\label{cc1}
 dC_j-\frac m2\,C_{j+1}dA-\frac m2\,C_{j-1}d\bar{A}=\frac m2\,\Phi\,
 C_{j+1}+\frac m2\,\Phi^* C_{j-1}+\Psi C_j,\eb
  \be\label{cc2}
 dC_j^{\,*}-\frac m2\,C_{j-1}^{\,*}dA-\frac m2\,C_{j+1}^{\,*}d\bar{A}=\frac m2\,\Phi\,
 C_{j-1}^{\,*}+\frac m2\,\Phi^{*} C_{j+1}^{\,*}+\Psi C_j^{\,*},\eb
 where $dA=(\delta_{\mu\nu}da_{\nu})_{\mu,\nu=1,\ldots,n}$ and
 $d\bar{A}=(\delta_{\mu\nu}d\bar{a}_{\nu})_{\mu,\nu=1,\ldots,n}$.

 Note that the expansion coefficients obey a set of algebraic
 relations. To derive them, let us first consider two multivalued
 solutions to Dirac equation, $u$ and $v$, that are square integrable
 at $|x|\rightarrow \infty$ and have the local expansions
 (\ref{fourier}) at the singularities. We shall assume that there
 exists a negative half-integer number $k_0$ such that
$$a^{(\nu)}_{\,k}(u)=b^{(\nu)}_{\,k}(u)=a^{(\nu)}_{\,k}(v)=b^{(\nu)}_{\,k}(v)=0,\qquad
\nu=1,\ldots,n,
$$
for all $k<k_0$. Using (\ref{stokes1}) and
 Stokes theorem,  calculate in two different ways the  integral
 $$\frac{m^2}{2}\!\!\!\!\!\!\!\!\int\limits_{\;\;\;\;\;\;\Cy\backslash\bigcup\limits_{\nu}
   D_{\varepsilon}(a_{\nu})}\!\!\!\!\!\!(\bar{u}_1v_1+\bar{u}_2v_2)\,
  idz\wedge d\bz= im\!\!\!\!\oint\limits_{\bigcup\limits_{\nu}
  \partial D_{\varepsilon}(a_{\nu})}\!\!\!\!\bar{u}_2v_1\,dz=-im\!\!\!\!\oint\limits_{\bigcup\limits_{\nu}
  \partial D_{\varepsilon}(a_{\nu})}\!\!\!\!\bar{u}_1v_2\,d\bz.$$
 Comparing the asymptotics of the corresponding boundary integrals as $\varepsilon\rightarrow0$,
 one obtains
 \be\label{auxx}
 \sum\limits_{\nu=1}^n\sum\limits_{k\in\Zb+\frac12}
 \left\{\overline{b^{(\nu)}_{\,k}(u)}\,a^{(\nu)}_{-k}(v)-
 \overline{a^{(\nu)}_{-k}(u)}\,b^{(\nu)}_{\,k}(v)\right\}(-1)^{k-1/2}\sin\pi\lambda_{\nu}=0.
 \eb

  If we now put $u=\tilde{\mathbf{w}}_{\mu}$,
 $v=\tilde{\mathbf{w}}_{\rho}$, then (\ref{auxx}) leads to the
 relation (analogous to (\ref{prelim}))
 $$\overline{b^{(\rho)}_{1/2}(\tilde{\mathbf{w}}_{\mu})}\,\sin\pi\lambda_{\rho}=
 b^{(\mu)}_{1/2}(\tilde{\mathbf{w}}_{\rho})\,\sin\pi\lambda_{\mu},$$
 or, in matrix notation,
 \be\label{alge1}C_{0}^{\,*}\,\sin\pi\Lambda=\left[\,\overline{C_{0}^{\,*}}\sin\pi\Lambda\right]^T,\qquad
 \Lambda=\left(\delta_{\mu\nu}\lambda_{\nu}\right)_{\mu,\nu=1,\ldots,n}.\eb
 On the other hand, the substitution
 $u=\tilde{\mathbf{w}}_{\mu}(\lambda)$,
 $v=\tilde{\mathbf{w}}_{\nu}^*(-\lambda)$ gives
\be\label{alge2}
 C_1(\lambda)\sin\pi\Lambda=\left[ C_1(-\lambda)\sin\pi\Lambda\right]^T.
 \eb

 Finally, observe that the entries of the $n$-dimensional vector
 $\partial_{\bz}\vec{\mathbf{w}}(\lambda)-\frac
 m2\,C_0^{\,*}(\lambda)\,\vec{\mathbf{w}}^* (-\lambda)$ belong to
 $\widetilde{\mathbf{W}}^{a,\lambda}$ and thus are all equal to
 zero. This gives two more relations,
 \be\label{alge3}
 C_0^{\,*}(\lambda)\overline{C_1(-\lambda)}=C_1^{\,*}(\lambda),\qquad
 C_0^{\,*}(\lambda)\overline{C_0^{\,*}(-\lambda)}=\mathbf{1}.
 \eb
 It is easy to see that for positive $\lambda_{\mu}$, $\lambda_{\nu}$
 both
 $\tilde{\mathbf{w}}_{\mu}(\lambda),\tilde{\mathbf{w}}_{\nu}(\lambda)\in\mathbf{W}^{a,\lambda}$, so we can
 calculate the inner product:
 $$\langle\tilde{\mathbf{w}}_{\mu}(\lambda),\tilde{\mathbf{w}}_{\nu}(\lambda)\rangle=
 -4\,\overline{b_{1/2}^{(\nu)}(\tilde{\mathbf{w}}_{\mu}(\lambda))}\,\sin\pi\lambda_{\nu}.$$
 This shows that the submatrix of $C_0^{\,*}(\lambda)\sin\pi\Lambda$,
 associated with the indices corresponding to positive
 $\{\lambda_{\rho}\}$, is negative definite. In a similar fashion, one
 finds
 $$\langle\tilde{\mathbf{w}}^*_{\mu}(-\lambda),\tilde{\mathbf{w}}^*_{\nu}(-\lambda)\rangle=
 4\overline{{b_{1/2}^{(\mu)}(\tilde{\mathbf{w}}_{\nu}(-\lambda))}}\,\sin\pi\lambda_{\mu}$$
 for negative $\lambda_{\mu}$, $\lambda_{\nu}$. Combining this formula
 with the second relation in (\ref{alge3}), one can prove that the
 submatrix of $\left(C_0^{\,*}(\lambda)\right)^{-1}\sin\pi\Lambda$,
 that corresponds to the ``negative'' indices, is positive definite.

 Let us return to deformation equations (\ref{cc1}) and
 (\ref{cc2}). In order to determine the unknown matrix-valued forms
 $\Phi$ and $\Phi^*$, let $j=-1$. One then finds
 \be\label{defs1}\Phi=-dA,\qquad \Phi^*=-C_0^{\,*}d\bar{A}\left(C_0^{\,*}\right)^{-1}.\eb
 Specializing to the case $j=0$, we calculate the form $\Psi$ and
 obtain a matrix equation,
 \be\label{defs2}
 \Psi=\frac m2 \,[dA,C_1]=dC_0^{\,*}\left(C_0^{\,*}\right)^{-1}+
 \frac m2\,\left[C_0^{\,*}d\bar{A}\left(C_0^{\,*}\right)^{-1},C_1^{\,*}\left(C_0^{\,*}\right)^{-1}\right].
 \eb
 For $j=1$, higher order coefficients arise. However, ``antiholomorphic'' part of (\ref{cc1}) and ``holomorphic'' part of
 (\ref{cc2}) comprise only the coefficients that are already involved:
 \be\label{defs3}
 d_{\bar{a}}C_1+\frac m2\,\left[C_0^{\,*}d\bar{A},\left(C_0^{\,*}\right)^{-1}\right]=0,
 \eb
 \be\label{defs4}
 d_aC_1^{\,*}+\frac m2\,[dA,C_0^{\,*}]-\frac m2\,[dA,C_1]\,C_1^{\,*}=0,
 \eb
 where $d_a=\sum\limits_{j=1}^n da_j\cdot\partial_{a_j}$ and
 $d_{\bar{a}}=\sum\limits_{j=1}^n d\bar{a}_j\cdot\partial_{\bar{a}_j}$
 In addition, the diagonal part of (\ref{cc1}) implies
 \be\label{defs5}
 d_a\,\mathrm{diag}\,C_1= \frac
 m2\,\mathrm{diag}\,\left([dA,C_1]\,C_1\right).\eb

 In order to write the deformation equations in more compact and
 standard form, introduce the notation
 \be\label{deffs1} G=C_0^{\,*}\sin\pi\Lambda,\qquad \Theta =\frac m2\,[dA,C_1],\qquad
 \Theta^{\dag}=\bar{\Theta}^T.\eb
 Using the symmetry relations (\ref{alge1})--(\ref{alge3}), one can
 show that  (\ref{defs2}) transforms into
 \be\label{deffs2} d G=\Theta G +G\Theta^{\dag}. \eb
 Instead of the equations (\ref{defs3}) and (\ref{defs4}) we have two
 conjugate relations
 \be\label{deffs3}d_{\bar{a}}\,C_1=\frac m2 \, [d\bar{A},G]\,G^{-1},\qquad d_{{a}}\,\overline{C}_1=\frac m2 \, [d{A},\overline{G}]\,{\overline{G}\,}^{-1},\eb
 and  the last equation (\ref{defs5}) can be
 rewritten as
 \be\label{deffs4}d_a\,\mathrm{diag}\,C_1=\mathrm{diag}\,(\Theta\,
 C_1).\eb

 We easily find from (\ref{deffs1}) and (\ref{deffs2})
 that $\mathrm{det}\,G=\mathrm{const}$. It is also very instructive to
 deduce the closedness of the 1-form
 $$ \Omega=\frac
 m2\sum\limits_{\nu=1}^n\left\{a^{(\nu)}_{1/2}(\tilde{\mathbf{w}}_{\nu}(\lambda))\,da_{\nu}+
 \overline{a^{(\nu)}_{1/2}(\tilde{\mathbf{w}}_{\nu}(-\lambda))}\;d\bar{a}_{\nu}\right\}=
  \frac m2\,\mathrm{Tr}\,(C_1dA+\overline{C}_1d\bar{A}), $$
 standing in the RHS of (\ref{dertau}), from the deformation equations. Indeed,
 $$ d\Omega=\frac m2\,\mathrm{Tr}\,\left(\Theta\, C_1\wedge dA+\frac
   m2\,[d\bar{A},G]\,G^{-1}\wedge dA+\overline{\Theta}\, \overline{C}_1\wedge d\bar{A}+\frac
   m2\,[dA,\overline{G}]\,\overline{G}^{-1}\wedge d\bar{A}\right)=$$
$$=-\frac{m^2}{4}\,\mathrm{Tr}\,\left(C_1dA\wedge
  C_1dA+\overline{C}_1d\bar{A}\wedge\overline{C}_1d\bar{A}+
 Gd\bar{A}\wedge G^{-1}dA+\overline{G}d{A}\wedge
 \overline{G}^{-1}d\bar{A}\right)=0, $$
 so the form $\Omega$ does represent the differential of a function.
\vspace{0,2cm}\\
 \textbf{Example}. As an illustration, let us find the explicit form
 of the deformation equations in the case $n=2$. Suppose that
 $\lambda_1>0$ and $\lambda_2<0$. Then $G_{11}<0$, $\det G<0$, and
 the matrix $G$ can be parametrized in the following way:
 $$ G=\chi \left(\begin{array}{cc}
 -e^{\eta}\sin\psi & e^{i\varphi}\cos \psi \\
 e^{-i\varphi}\cos \psi & e^{-\eta}\sin\psi\end{array}\right),\qquad \chi,\eta,\psi,\varphi\in\Rb,$$
 where $0<\psi<\pi$, $\chi>0$. We shall also denote
 $$ C_1=\left(\begin{array}{cc}
\Lambda_{11} & \Lambda_{12} \\ \Lambda_{21} &
\Lambda_{22}\end{array}\right), \qquad
q=m(a_2-a_1)/2,\quad\bar{q}=m(\bar{a}_2-\bar{a}_1)/2.$$
 From (\ref{deffs2}) one obtains
$$ \frac{\partial G}{\partial q}=- {\chi}^{-1}\left(\begin{array}{cc}
 \Lambda_{12}e^{-i\varphi}\cos\psi & \Lambda_{12}e^{-\eta}\sin \psi \\
 \Lambda_{21}e^{\eta}\sin \psi  &
 -\Lambda_{21}e^{i\varphi}\cos\psi\end{array}\right),$$ $$
 \frac{\partial G}{\partial \bar{q}}=- {\chi}^{-1}\left(\begin{array}{cc}
 \overline{\Lambda}_{12}e^{i\varphi}\cos\psi & \overline{\Lambda}_{21}e^{\eta}\sin \psi \\
 \overline{\Lambda}_{12}e^{-\eta}\sin \psi  &
 -\overline{\Lambda}_{21}e^{-i\varphi}\cos\psi\end{array}\right).$$
 This leads to the relations
$$\partial_q\varphi=i\,\mathrm{tg}^2\psi\,\partial_q\eta,\qquad
\partial_{\bar{q}}\varphi=-i\,\mathrm{tg}^2\psi\,\partial_{\bar{q}}\eta,$$
$$\Lambda_{12}=e^{\eta+i\varphi}\left(\partial_q\psi-i\,\mathrm{ctg}\,\psi\,\partial_q\varphi\right),\qquad
\Lambda_{21}=e^{-\eta-i\varphi}\left(\partial_q\psi+i\,\mathrm{ctg}\,\psi\,\partial_q\varphi\right).$$
 Next, the first formula in (\ref{deffs3}) implies that
 \be\label{exf}\frac{\partial C_1}{\partial\bar{q}}=-\left(\begin{array}{cc}
\cos^2\psi & e^{\eta+i\varphi}\cos\psi\sin\psi \\
e^{-\eta-i\varphi}\cos\psi\sin\psi & -\cos^2\psi \end{array}\right).\eb
 Off-diagonal part of this relation leads to a system of coupled
 differential equations,
 $$\left\{\begin{array}{l}\partial_{q\bar{q}}\psi+{\frac{\cos\psi}{\sin^3\psi}}\,\partial_q\varphi\,\partial_{\bar{q}}\varphi+
 \sin\psi\cos\psi=0, \\
 \partial_{q\bar{q}}\varphi=\frac{1}{\sin\psi\cos\psi}\,\left(
 \partial_q\varphi\,\partial_{\bar{q}}\psi+\partial_{\bar{q}}\varphi\,\partial_q\psi\right).\end{array}\right.
 $$
 Finally, the formula (\ref{deffs4}) and the diagonal part of (\ref{exf})
 give the second logarithmic derivatives of the $\tau$-function:
 \be\label{exxf}
 \left\{ \begin{array}{l}
 \partial_{q\bar{q}}\ln\tau=\cos^2\psi ,\\
 \partial_{qq}\ln\tau=(\partial_q \psi)^2+\mathrm{ctg}^2\psi\,(\partial_q
 \varphi)^2 ,\\
 \partial_{\bar{q}\bar{q}}\ln\tau=(\partial_{\bar{q}} \psi)^2+\mathrm{ctg}^2\psi\,(\partial_{\bar{q}} \varphi)^2 .
 \end{array}
 \right.
 \eb
 \section{Discussion}
 When one tries to generalize the above theory, a natural question
 arises: if it is possible to develop the theory of monodromy
 preserving deformations for the massive Dirac operator on the
 arbitrary two-dimensional surface $M$ with a metric? It appears that
 $M$ should be then a homogeneous space for a group $G$, acting on $M$ by
 isometries. There are only five such surfaces: plane ($G=E(2)$),
 cylinder and torus ($G=T^2$), Poincar\'e disk ($G=PSU(1,1)$)
 and the sphere ($G=PSU(2)$). The plane and hyperbolic disk were
 studied earlier by different authors (see references in the
 Introduction). The present paper is devoted to the
 cylindrical geometry. It is interesting to note that the derivation of all ``implicit''
 results (factorized form of the derivatives of Green functions,
 deformation equations, etc.) can be transferred almost literally to
 the case of torus. What is even more important --- the technical
 results, obtained in this work (namely, the formulae for the
 one-point projections in the Theorem~\ref{proj}) allow to calculate
 the $\tau$-functions on the torus explicitly. I hope to discuss these
 matters in greater detail elsewhere.

 The second task is to give a proper formulation and solution of the
 problem in the quantum field theory language. Let us interpret the
 coordinate along the cylinder axis as time, with the space coordinate
 living on the circle. The time axis is splitted by the branchcuts
 $b_1,\ldots,b_n$ into $n+1$ intervals. The evolution in each interval
 is governed by the Dirac hamiltonian, which is diagonalized in the
 free-fermion basis. These free fermions, however, obey different
 periodicity conditions (statistics) in different
 intervals. Corresponding Fock spaces are intertwined by the
 monodromy fields, whose correlation functions can be written in terms
 of Lehmann expansion over intermediate eigenstates of the
 hamiltonians. (In the two-point case, this corresponds to the
 expansion of the determinant (\ref{taun2})). The problem transforms
   then into  the calculation of form factors of monodromy fields in
   the finite volume.

 Another important problem is the investigation of the ultraviolet
 ($m\rightarrow0$) asymptotics of the $\tau$-functions. On the plane,
 the connection between the Ising model and singular Dirac operators
 was already used in \cite{short} to give a rigorous proof of the
 Luther-Peschel formula.
 \\
\textbf{Acknowledgements}. I would like to thank A.~I.~Bugrij and
V.~N.~Roubtsov for constant support and numerous stimulating
discussions. I am grateful to S.~Pakuliak for his lectures on the
infinite-dimensional grassmannians and boson-fermion
correspondence. I would also like to express my gratitude to
J.~Palmer, whose clear ideas made this work possible.
 
\end{document}